\documentclass{article}
\usepackage[utf8]{inputenc}
\usepackage{graphicx}
\usepackage{amsmath}
\usepackage{float}
\usepackage{cleveref}
\usepackage{caption}
\usepackage{subcaption}
\usepackage{booktabs} 
\usepackage{xcolor}
\usepackage{soul}
\usepackage[affil-it]{authblk}
\usepackage[numbers]{natbib}
	
\usepackage[euler]{textgreek}

\usepackage{mymacros_11Sep20}

\graphicspath{{./}{./Figures/}}

\crefname{equation}{Equation}{Equations}
\Crefname{equation}{Equation}{Equations}
\crefname{table}{Table}{Tables}
\Crefname{table}{Table}{Tables}

\crefname{figure}{Figure}{Figures}
\Crefname{figure}{Figure}{Figures}
\title{Constant Chemical Potential-Quantum Mechanical-Molecular Dynamics simulations of the Graphene-electrolyte double layer  }
\author[1]{Nicodemo Di Pasquale  \thanks{Corresponding author: nicodemo.dipasquale@brunel.ac.uk}}
\author[2]{Aaron R. Finney}
\author[3,4]{Joshua Elliott}
\author[3]{Paola Carbone} 
\author[2]{Matteo Salvalaglio}

\affil[1]{Department of Chemical Engineering, Brunel University London, Uxbridge, UB8 3PH, United Kingdom}
\affil[2]{Department of Chemical Engineering, University College London, London, WC1E 7JE, United Kingdom}
\affil[3]{Department of Chemical Engineering, University of Manchester,  Manchester, M13 9PL, United Kingdom}
\affil[4]{Diamond Light Source, Harwell Science and Innovation Park, Didcot, Oxfordshire OX11 8UQ, United Kingdom}

\date{October 2021}

\begin{document}

\maketitle

\begin{abstract}
  \noindent We present the coupling of two frameworks---the pseudo-open boundary simulation method known as constant potential Molecular Dynamics simulations (C$\mu$MD), combined with QMMD calculations---to describe the properties of graphene electrodes in contact with electrolytes. 
  
  The resulting C$\mu$QMMD model was then applied to three ionic solutions (LiCl, NaCl and KCl in water) at bulk solution concentrations ranging from 0.5 M up to 6 M in contact with a charged graphene electrode. The new approach we are describing here provides a simulation protocol to control the concentration of the electrolyte solutions while including the effects of a fully polarizable electrode surface.
   Thanks to this coupling, we are able to accurately model both the electrode and solution side of the double layer and provide a thorough analysis of the properties of electrolytes at charged interfaces, such as the screening ability of the electrolyte and the electrostatic potential profile. We also report the calculation of the integral electrochemical double layer capacitance in the whole range of concentrations analysed for each ionic species, while the QM simulations provide access to the differential and integral quantum capacitance. We highlight how subtle features, such as the adsorption of potassium at the interface or the tendency of the ions to form clusters, emerge from our simulations, contribute to explaining the ability of graphene to store charge and suggest implications for desalination.
\end{abstract}

\section{Introduction}

Interest in graphene-based devices has grown in recent years, thanks of the versatility and physical characteristics of this new material, in particular for applications in which it is in contact with an electrolyte solution. Use of nanoporous graphene as a membrane for water desalination \citep{Cohen-Tanugi2012,Heiranian2021} is one important example. The presence of pores of equal size to the electrolytes allows the selective passage of water through the membrane. Combined with the atomic scale thickness of graphene, this can lead to the creation of desalination membranes with higher performances than common polymer-based ones \citep{Surwade2015}. Another promising technologically relevant applications is the use of graphene electrodes in electrochemical double layer (super)capacitor (EDLC) devices\citep{Simon2008, Wang2021, Elliott2022rev}.  In fact, graphene \citep{Wang2009capacitors,Liu2010,Yu2010,Zhang2010}, porous activated carbon \citep{Zhu2011}  and carbon nanotube \citep{An2001,Yang2019} electrodes potentially have relatively high charge storage capacity and a favourable specific energy to power ratio, due to rapid charge-discharge cycling \citep{Liu2010} controlled by changes of an applied potential, together with lifetimes that can reach millions of cycles \citep{Zhu2011}.

Typically, charge storage at carbonaceous electrodes is a non-faradaic process, where mobile ionic species accumulate at the interface between the electrode and the liquid phase.
 An important class of systems of this kind, which has gained lots of attention recently, is represented by cheap and easy-to-prepare aqueous-based electrolytes in contact with a graphene electrode \citep{Elliott2022rev}. 
  Carbon-based EDLCs with aqueous-based electrolytes do not generally suffer from electrochemical degradation, can be non-toxic,  and provide an attractive alternative solution at the problem of energy storage compared with traditional battery devices. Combined with a longer lifetime and high power density,\citep{Merlet2013} these energy storage systems could be increasingly applied to power small electronic devices and for acceleration and breaking in electrical vehicles \cite{Wang2021}.

Several experimental works were undertaken to understand the physicochemical properties of neutral and charged graphene interfaces in contact with electrolyte solutions and the nature of these systems charge storage capacity \citep{Iamprasertkun2019,Yang2017,Qu2008}. However, the delicate balance between hydration-free energy and surface effects, which regulate the physisorption of ionic species at surfaces, resulted in conflicting experimental findings (see \citep{Elliott2020} for a more detailed account). For instance, there are reports both supporting the conclusion that the capacitance of graphene films is ion-independent \citep{Yang2017}, as well as contrasting observations suggesting that basal capacitance is instead ion-specific (with, for example, a greater propensity for Na$^+$ and K$^+$ adsorption over Li$^+$ adsorption at negatively charged electrodes in the case of group I cations)\citep{Qu2008}. Atomic-scale defects in the graphitic surface, its topography, dimensionality and chemical modifications are difficult to control and have non-negligible effects in experimental measurements. As an example, mechanical cutting produces structural defects known as ``dangling bonds'' which modifies the measured capacitance of the sample \citep{Iamprasertkun2019,Wang2008}. In this respect, a model of the graphene interface and its interactions with an electrolyte solution can exclude all the spurious effects coming from uncontrolled defects and chemical modification of the surface. Molecular modelling and simulations can help to improve  understanding of the mechanisms involved in such complex systems and guide the interpretation of experimental results.

Many key features of supercapacitive devices are underpinned by the properties of the electrochemical double layer, and their responses to the charging of the electrode. Gouy-Chapman theory \citep{Gouy1910,Chapman1913} describes the double layer as a diffuse charged layer in the solution that compensates an applied surface charge on the electrode. Modifications to this model include the adsorption of counter-ions at the surface in the so-called Stern layer \citep{Stern1924}.
The development of a mean-field theory based on the Poisson-Boltzmann lattice-gas model \citep{Popovic2013} has shown that features not present in the Gouy-Chapman theory, such as steric effects, ion correlations, and preferential adsorption \citep{Borukhov1997,Fedorov2008,Howard2010} need to be accounted for in order to correctly describe the interactions between the ions and the electrode. 
Mechanistic insight for these kinds of effects and how they control charge storage can be gained by atomistic simulations of the graphene/electrolyte interface; these also enable the evaluation of ensemble properties, such as the free energy of adsorption of the ions at the interface \citep{Misra2021}. Furthermore, simulations can establish the effect of solution concentration on ion accumulation at the electrode, their interfacial structure, and dynamical properties. 

In order to compare simulations with a macroscopic system, this adsorption should ideally be modelled in the presence of bulk electroneutral solution with fixed composition to ensure a constant driving force for the adsorption at a charged surface.
This can be obtained for example as shown in \citet{Finney2021}, where the authors performed MD simulations using constant chemical potential MD simulations, C$\mu$MD \citep{Perego2015}, which mimics open-boundary conditions. With C$\mu$MD, the authors simulated NaCl(aq) with concentrations spanning $\sim 0.1-10$~M at graphite surfaces. Their results indicate that the interface charge screening behaviour is a function of bulk solution concentration, with a transition (at $\sim 1$M) from diffuse charge screening, qualitatively consistent with the picture from simple mean field models, to a complex multi-layered structuring that systematically either over or under screens the surface potential. The multiple charged layers result from ion finite-size effects, over-compensation of the surface charge by oppositely charged ions closest to the surface, and non-idealities in solution, i.e., when the hypothesis of non-interaction between oppositely charged ions breaks down for large ions concentrations \citep{Doblhoff-Dier2021}. This last effect also has  consequences on the conductance of the ions, which deviates from the prediction of the Nerst-Einstein equations \citep{France-Lanord2019}.

Together with a constant driving force for ion adsorption from the bulk,  another important effect to consider in the description of such systems is the the polarisation of the electrode exerted by the adsorbing electrolytes \citep{Elliott2020}. 
Classical simulations typically model the non-bonded interactions between atoms within the electrolyte and atoms belonging to the interface using additive pairwise potentials such as the Lennard-Jones potential and Coulomb interactions between fixed point atom charges.
Polarisation can be introduced using e.g., oscillating charge models, or by fitting short-range potentials to binding energies obtained from \textit{ab initio} methods \citep{Misra2021,Misra2021b,Williams2017}. However, these models may not accurately capture the complex many-body effect associated with charge polarization at the electrode-solution interface. Another way to include polarisation in classical MD simulations is through the constant potential method developed in \citep{Siepmann1995}. This constant potential method has been successfully deployed to describe the properties of the electrochemical double layer of aqueous electrolytes and ionic liquids in contact with metal electrodes such as Au and Cu. Despite its successes, one of the key approximations of the constant potential method is that the electrode is fully metallic and can perfectly screen charges, which is not the case for (semimetallic) graphene \citep{Elliott2020}.

On the other hand, a full Quantum Mechanical (QM) treatment of the interactions between the electrolyte and the substrate is still unfeasible, due to the length (tens of nm) and time (hundreds of ns) scales required for modeling the effect of the aqueous electrolytes. 
However, while the full QM model of the electrode/electrolyte system is out of reach, QM calculations can be used to compute a set of atomic partial charges on the electrode in the presence of the electrostatic potential arising from the position of the electrolyte atoms.
This is exactly the spirit of our QMMD scheme, where QM calculations are coupled to MD simulations at fixed intervals of time integration. As such, the surface atom partial charges within the classical force field are updated on the fly.
In a more recent development Machine Learning models have recently proven to be a viable option in tuning the surface polarization if the scope of the system becomes too large for QM simulations. This is achieved by replacing the QM calculations with a Neural Network (NN) model trained to reproduce results from a wide range of QM calculations with varying distributions of electrolytes in solution. The NN acts as a polarizable-like force field, combining fast classical MD simulations with more accurate QM calculations of the interface polarization \citep{DiPasquale2020}.

%In particular, it is known that in Ionic Liquids with a more complex molecular structure the structure of the electrical double layer is radically different from the EDL in the case of electrolyte systems \citep{Fedorov2014}.

This present work leverages the QMMD framework introduced in \citep{Elliott2020} and the C$\mu$MD introduced in \citep{Perego2015,Finney2021}. The approach simultaneously captures surface polarization and concentration effects that can modify the structure and composition of the electrochemical double layer. 
We use the resulting C$\mu$QMMD protocol to examine interfaces between aqueous alkali chloride solutions at different concentrations with a graphene electrode surface, elucidating complex interfacial structure, dynamics, and electrochemical properties. 

This paper is organized as follows: we first provide a brief overview of the QMMD and C$\mu$MD protocols, pointing to the relevant literature for the interested reader; we present the systems to which we apply the C$\mu$QMMD framework: a charged graphene electrode in contact with three different electrolyte solution, NaCl(aq), LiCl(aq), KCl(aq) at different concentrations. We derive the electrical properties of the interface in terms of the screening factor and electrical potential and calculate the total integral capacitance of this system by deriving the quantum and electrical double layer capacitance. Finally, we discuss the effects of complex solute speciation on the performance of graphene-electrolyte devices and draw some conclusions regarding this new proposed simulation scheme.

\section{Computational Models}

% Motivation for QMMD
In order to capture the dynamic polarization of a charged graphene surface in response to the evolving configuration of an electrolyte at a prescribed concentration, we coupled the classical C$\mu$MD simulation to the electronic structure theory calculations at regular time intervals. We will give a more detailed account of both models (C$\mu$MD and QMMD) in the following sections, while here we will only discuss their coupling.

\begin{figure}[ht]
    \centering
    \includegraphics[width=.65\textwidth]{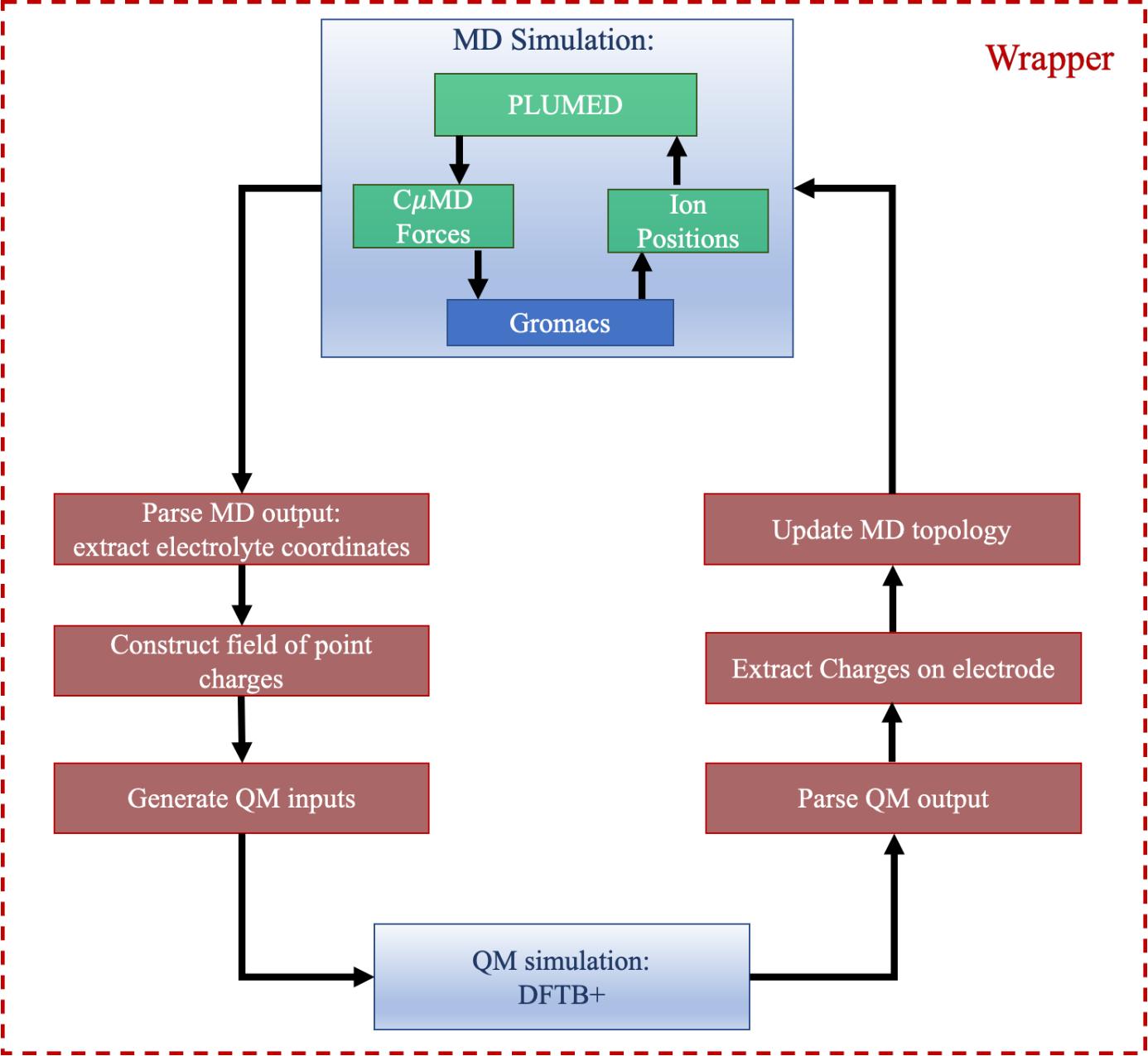}
    \caption{The computational workflow adopted in this work highlighting the two ``black boxes'' (the MD software and the QM software) in the blue squares and the operations included in the python wrapper (red squares).  }
    \label{fig:loop}
\end{figure}

A sketch of the sequence of the operations involved  is given in \cref{fig:loop}. All the operations shown in \cref{fig:loop} are obtained through an in-house python wrapper. During the MD time integration obtained with GROMACS 2018.4 MD package  \citep{Abraham2015}, ion positions are passed to the Plumed software (v. 2.7)  \cite{Tribello2014} patched with GROMACS, to compute the C$\mu$MD forces (see \cref{sec:cmu} for more details). After the evolution of the atom positions, the final configuration of the electrolyte is extracted to compute the electrostatic potential. In turn, this latter quantity is used as input for the QM calculations obtained with the \textsc{Dftb+} software package \citep{Hourahine2020}. From the QM results, the distribution of the charges on the graphene is extracted (see \cref{sec:qmmd} for more details) and used as input for the new iteration  of the loop.

%using the C$\mu$MD model implemented in PLUMED  . 

% Pseudo alogorithm for QMMD

\subsection{C$\mu$MD Model} \label{sec:cmu}
\begin{figure}[ht]
    \centering
    \includegraphics[width=.95\textwidth]{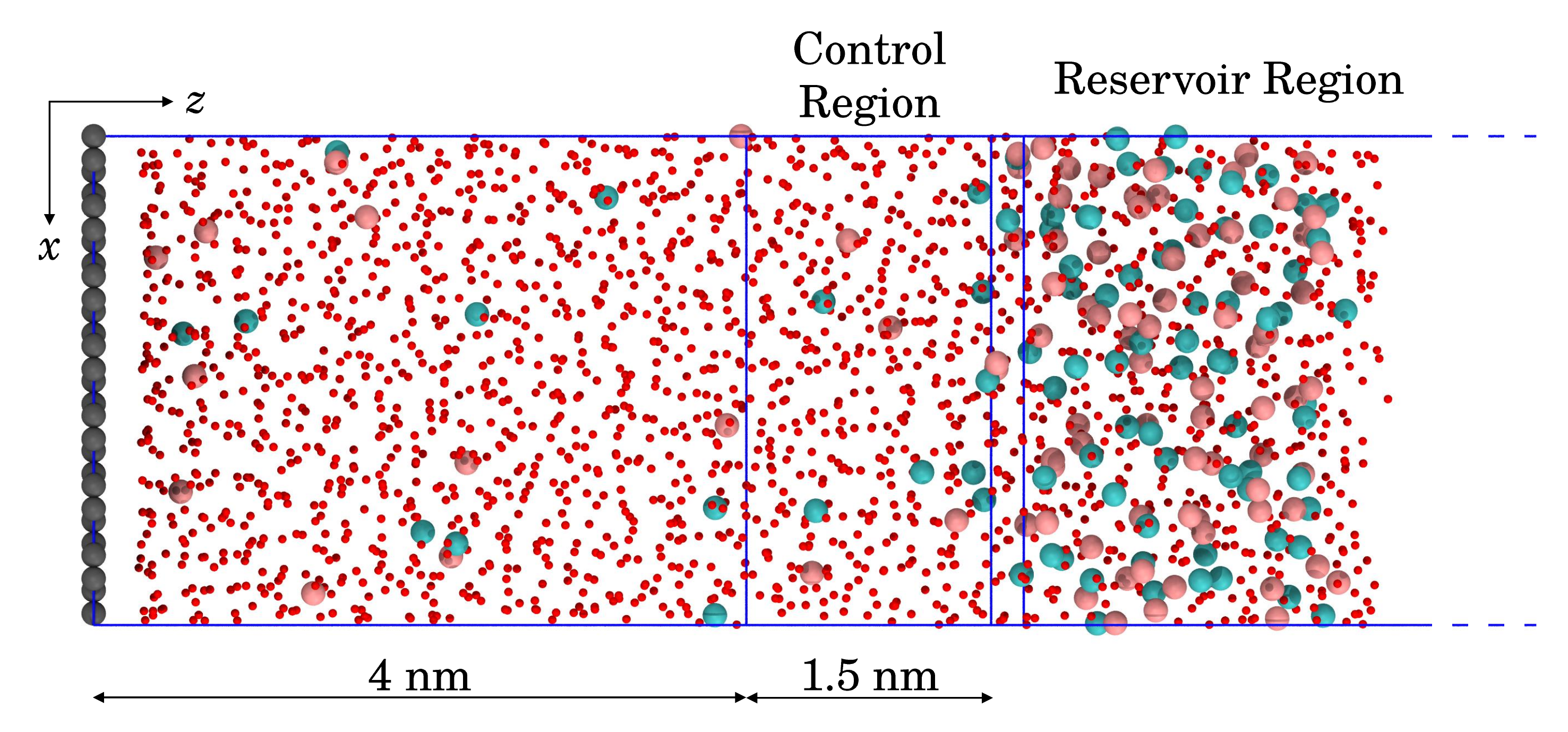}
    \caption{Example configuration from a C$\mu$QMMD simulation of KCl(aq) in contact with graphene in this work projected onto simulation $x,z$ dimensions. K$^+$, Cl$^-$, O of water and C of graphene are shown by the pink, cyan, red and grey spheres. The blue lines highlight the C$\mu$QMMD control and reservoir regions, which also indicate the simulation cell boundaries. An extended vacuum region, around 8 nm in $z$, is truncated in the image.}
    \label{fig:snap}
\end{figure}

\noindent The graphene electrode we considered is located at $z=0$ and is in contact with an electrolyte slab of thickness 8 nm. A further 8 nm of vacuum separates the system from its periodically repeating images. The electrolyte phase is divided into three regions: the first region starts at the graphene electrode up to a distance of 4 nm. The second one is the \textit{control region}, which is used to control the concentrations. The third region is the \textit{reservoir region} which provides the reservoir of ions to adjust the concentration of the electrolytes in the other regions.  \Cref{fig:snap} provides an example of the set-up adopted in this work, where we highlighted the different  C$\mu$MD simulation cell regions. 

The control of the concentration of the ions in solution is obtained by applying a force at the edge of the reservoir region according to a continuous function of the form,
\begin{equation}
    F_i^\mu(z) = k_i(n_i^\mathrm{CR}-n_i^0) \left[\frac{1}{4 \omega} \left( 1 + \mathrm{cosh} \left( \frac{z-z_F}{\omega}
    \right) \right)^{-1} \right].
\end{equation}
Here, $\omega$ was set to 0.2 nm, and represents the width of the force region (between the control and reservoir regions highlighted by the blue lines in \cref{fig:snap}) while $k$ was $2 \times 10^4$ kJ mol$^{-1}$ nm$^{-1}$,  giving the correct densities in the bulk (see \citep{Finney2021} for a discussion on these parameters).
$n^0$ is the target ion number density, while $n^{\mathrm{CR}}$ is the density calculated instantaneously during time integration in the control region.
Finally, $z_F$ is the position in $z$ where the C$\mu$MD forces are applied. In our simulations this is set to 5.5 nm beyond the graphene surface.
Using this approach, the densities of cations and anions are constrained in the control region to maintain target concentrations of 0.5, 2.0, 3.0, 4.0, 4.4 and 6 M.
At each MD time-step ion positions are passed to Plumed in order to compute the C$\mu$MD forces only acting on those ions in the region of $z_F$. No external forces are applied to the ions outside of this region, and any local change in the ion density at the interface results from the physical interactions between graphene and the solution.

\subsection{QMMD Model} \label{sec:qmmd}

The generality of  electronic structure theory and its ability to reproduce the electronic charge density distribution in semiconductors, metals, and semimetals implies that that the QMMD approach can describe both long- and short-ranged redistribution of the surface charge induced by the presence of the electrolyte.
 Within each iteration (see \cref{fig:loop}) of our scheme, the fully classical system is taken as input for a quantum mechanical calculation. The simulation box is partitioned into surface atoms whose electronic structure is explicitly treated, and electrolyte atoms that are converted into a set of point charges. The point charges take the values of the partial charges contained in the classical force field and form the background electrostatic potential during the computation of the electron structure.
Upon derivation of the electronic structure, partitioning of the charge density via Mulliken population analysis yields the set of surface atom partial charges, which are then passed to the classical force field. Finally, a short MD trajectory on the order of several picoseconds can then be carried out (in the presence of the quantum mechanically polarized surface) to generate the electrolyte configuration for the following iteration. In our simulations we employ a coupling between QM and MD calculations of 5 ps. We previously found for this class of systems that 5 ps represents a good compromise in terms of computational accuracy of the computed charges (0.004 $e$) vs computing time when compared with a QMMD simulation where the charges were updated at every MD time step \cite{Elliott2020}.

% QMMD in practice DFTB+/ Gromacs 
In practice, in order to describe the electronic structure of solid-electrolyte interfaces on the length scales required, we leverage the self-consistent charge Density Functional Tight-Binding (SCC-DFTB) \citep{Elstner1998} approach, which is an approximation to Kohn-Sham Density Functional Theory. 
The empirical description in our \textsc{Dftb+} calculations of the interactions between the C atoms in the surface are described by the mio-1-1 parameter set. The SCC charge threshold and Fermi temperature have been set to $1\times10^{-2}$ Hartree and 300 K, respectively.
Whereas, on first inspection, these criteria can be considered loose and should not be adopted for the calculation of the total electronic energy, rigorous testing in our previous works \citep{Elliott2020,Elliott2022} found that they provide a sufficiently accurate description of the surface charge distribution with respect to fully converged simulations, at a fraction of the computational cost.  
Finally, to compute the partial charges passed to the graphene force field at each MD step, we perform a Mulliken population analysis \citep{Mulliken1955}, which gives reasonable results for this class of systems\citep{Elliott2020,Elliott2022}.

%\hl{This scheme does not include charge transfer. However, when used with elements for which the nature of the interactions between electrolytes and graphene carbons is mainly ionic there is no need to consider charge transfer. In this work we consider alkali-earth-metal ions for which it was shown that the nature of the interactions is mainly ionic} \citep{Zhou2020}.

\subsection{Simulations Details}

In our simulations, we consider a graphene electrode composed of 336 carbon atoms in contact with aqueous electrolyte solutions. We investigated three electrolyte systems, NaCl, KCl and LiCl at concentrations ranging from 0.5 M to 6 M. However, due to the solubility limits of the KCl(aq)  \citep{Haynes2016,Zeron2019}, we limit the investigated concentrations to 4.4 M for the KCl system. 

Our simulations are carried out at constant surface charge, which makes it difficult to draw comparisons across different electrodes since the potential applied is not necessarily constant. As such, when we compute the capacitance, we use the potential drop of the neutral electrode as a reference. This approach has been applied previously to compare the properties of the electrochemical double layer for different electrolytes \citep{Ho2013}.
Each operating condition was therefore repeated for two different total charges of the electrode: a charged graphene layer with a constant charge on the surface \citep{Xu2020} $\sigma$ of -0.449 e nm$^{-2}$ (-0.0719 C/$m^2$)  and a neutral one ($\sigma = 0$).  %However, the surface charge density is not uniformly distributed, and it is modulated by the electrolyte charge distribution via a feedback loop obtained by frequently coupling the MD engine with QM calculations, as detailed in the next sections. We focus on the negatively charged electrode as a proof of concept for the method. 

Structural analyses of the solutions are carried out using PLUMED by post-processing the simulation trajectories. The first-shell coordination numbers for cations with anions ($N_{\mathrm{X-Cl}}$) and cations with water oxygen atoms ($N_{\mathrm{X-Ow}}$) were computed using a continuous switching function:
\begin{equation}\label{eq:sphere}
    N = \frac{1}{M} \sum_i^M \exp{\cip{ \frac{-(r-d_0)^2}{2r_0^2}}}
\end{equation}
where $r$ are distances between atoms, $r_0$ was 0.01 nm, and $d_0$ was chosen such that the function goes smoothly from one to zero at the position of the first minimum in radial distribution functions for the cations with anions and water oxygen atoms.
This ensured that a conservative definition of first-shell coordination was adopted in the analyses. %Averages are computed for all $M$ cations.
Coordination numbers were evaluated in 1.3 nm regions in $z$ closest to the graphene surface and 3.5 nm from the surface, representing the double layer and bulk solution regions, respectively.
The first coordination sphere distributions for ions were used to construct a graph of ion-ion contacts using the NetworkX Python library \cite{Hagberg2008}. This allowed us to identify and compute the size of the ion clusters formed. Ion clusters at the interface and within the bulk were identified by sampling the regions defined for computing the coordination numbers.

%In this regard, we extended the python wrapper we presented in our previous publication (see \citep{Elliott2020} and next paragraph) to incorporate C$\mu$MD model implemented in PLUMED \cite{Tribello2014}. 

Molecular dynamics calculations in the NVT ensemble are carried out using GROMACS \citep{Berendsen1995,Spoel2005}, version 2018.4. The leapfrog algorithm with a timestep of 1 fs was used to integrate the equations of motion at a constant temperature of 298.15 K, controlled with the Nosé-Hoover thermostat, with a relaxation time of 0.1 ps. Long-range electrostatic interactions were treated using the particle-mesh Ewald approach, with a cut-off of 1.4 nm. Non-bonded interactions were computed using a Lennard-Jones 12-6 potential, truncated smoothly at 1.0 nm using a switch function starting at a distance of 0.99 nm. In all simulations, graphene carbon atoms were frozen, and water was modelled using the SPC/E model \citep{Berendsen1987} with the SETTLE algorithm used to maintain rigid molecule geometries \citep{Miyamoto1992}. This choice is compatible with the Werder water-graphene parameters that reproduce the experimentally measured graphene/water contact angle \citep{Wang2009capacitors, Huang2012}.
Ion force field parameters (for K$^+$, Li$^+$, Na$^+$, Cl$^+$), also compatible with the SPC/E model, are taken from the work of \citet{Joung2008}. In order to prevent water molecules and ions from escaping the solution into the vacuum space, we added a fixed wall above the reservoir, interacting with water molecules and ions only through a short-range Lennard-Jones potential.

We equilibrated each system for 20~ns followed by 130~ns production runs to collect data for subsequent analyses of the steady-state structure of the interface. In all analyses discussed below, mean values and standard deviations (error bars) are obtained via averaging performed using 5~ns windows.

\section{Results and Discussion}

Thanks to the simulation protocol implemented, electroneutral solutions with fixed ion concentrations can be maintained in the Control Region in Figure \ref{fig:snap}, representing bulk solutions in equilibrium with the electrode-solution interfaces. This allows us to compare the behaviour of different electrolytes while controlling the electrolyte background concentration. 

\subsection{Density Profiles}

\begin{figure*}
  \begin{subfigure}[b]{0.33\textwidth}
   \centering
    \includegraphics[width=\textwidth]{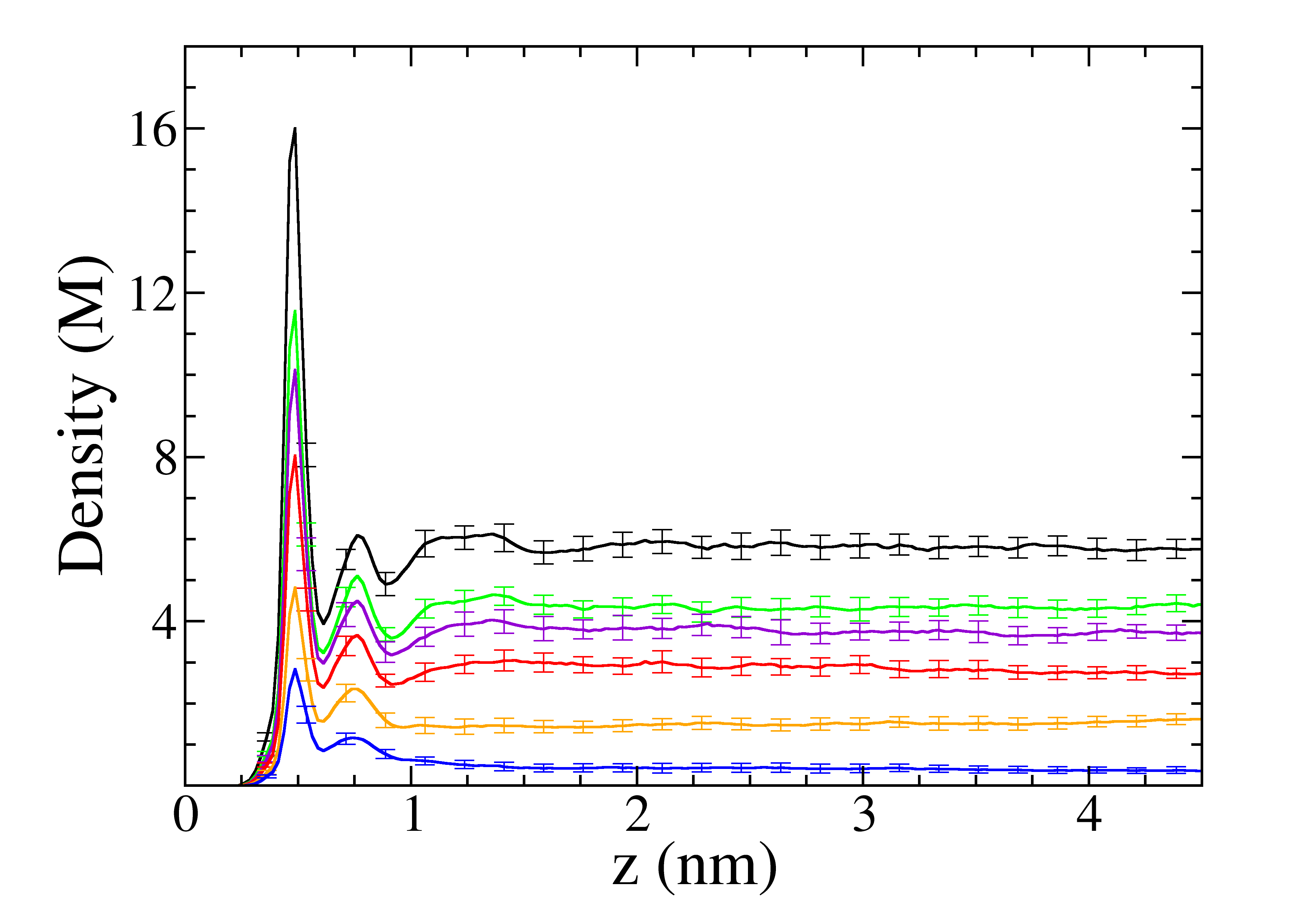}
    \caption{ Na$^+$} 
    \label{fig:na} 
    \vspace{2ex}
  \end{subfigure}%% 
 \begin{subfigure}[b]{0.33\textwidth}
    \centering
   \includegraphics[width=\textwidth]{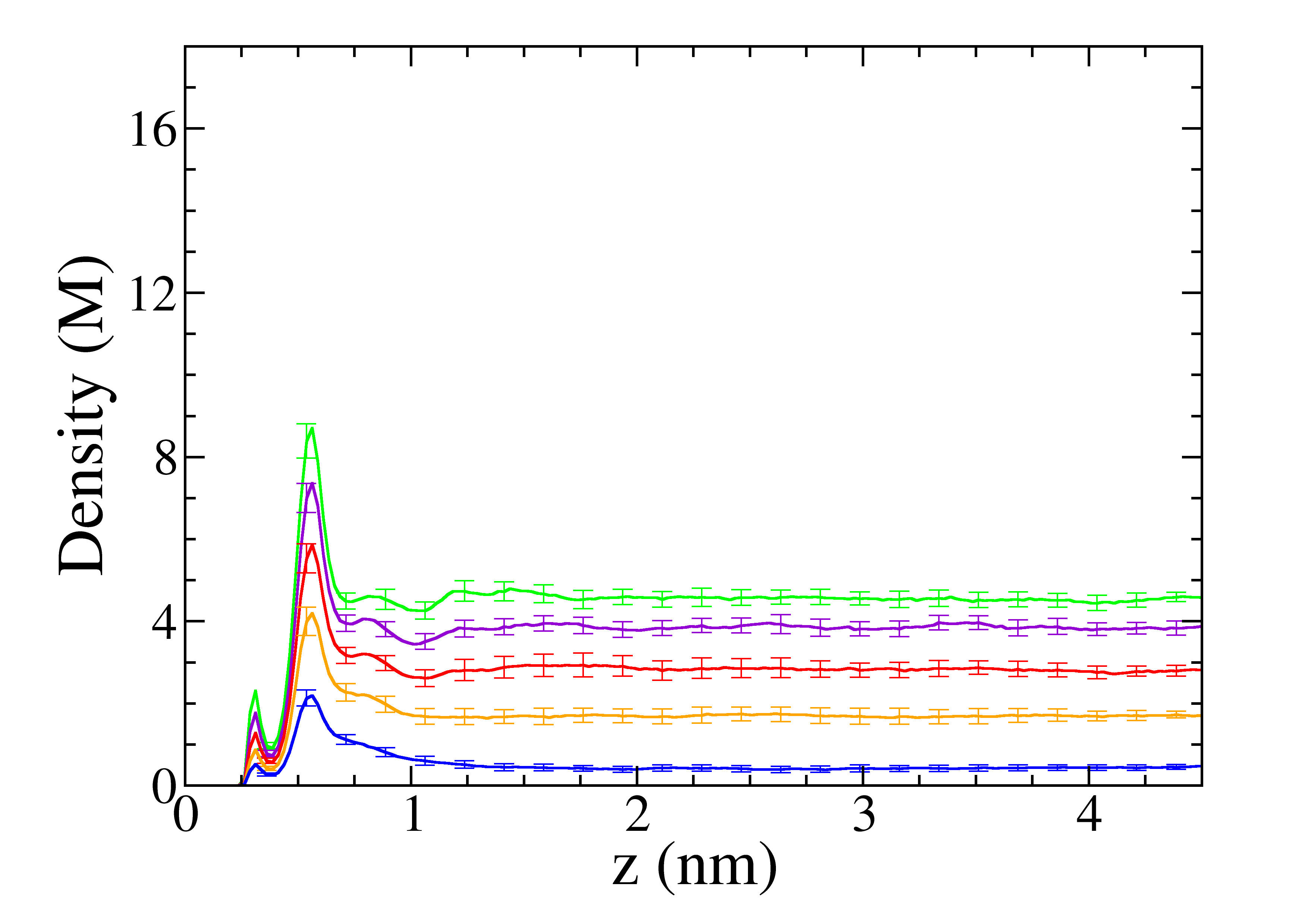}
    \caption{ K$^+$} 
     \label{fig:k} 
    \vspace{2ex}
  \end{subfigure} 
    \begin{subfigure}[b]{0.33\textwidth}
   \centering
    \includegraphics[width=\textwidth]{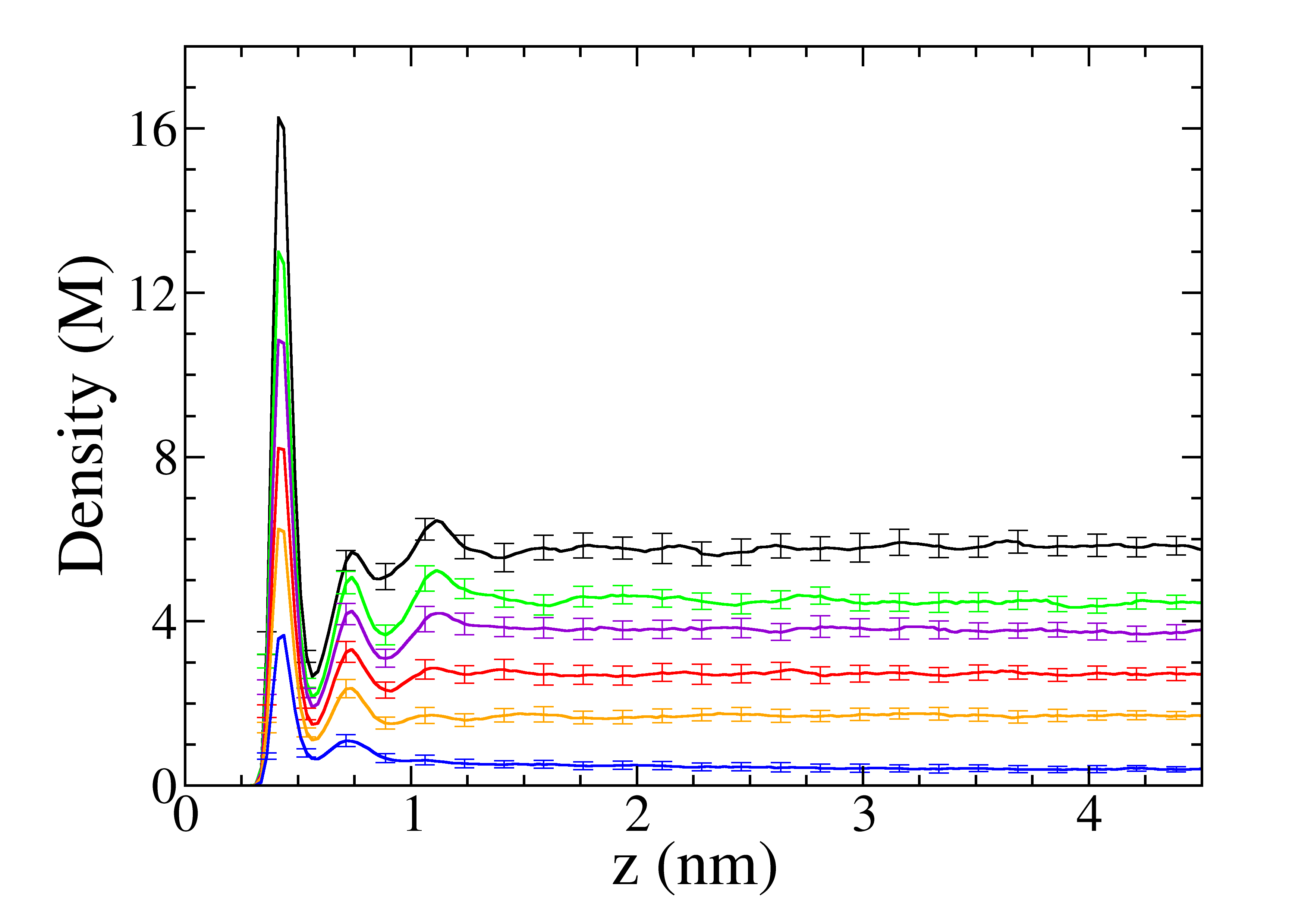}
    \caption{ Li$^+$} 
    \label{fig:li} 
    \vspace{2ex}
  \end{subfigure}%% 
  \\
    \begin{subfigure}[b]{0.33\textwidth}
   \centering
    \includegraphics[width=\textwidth]{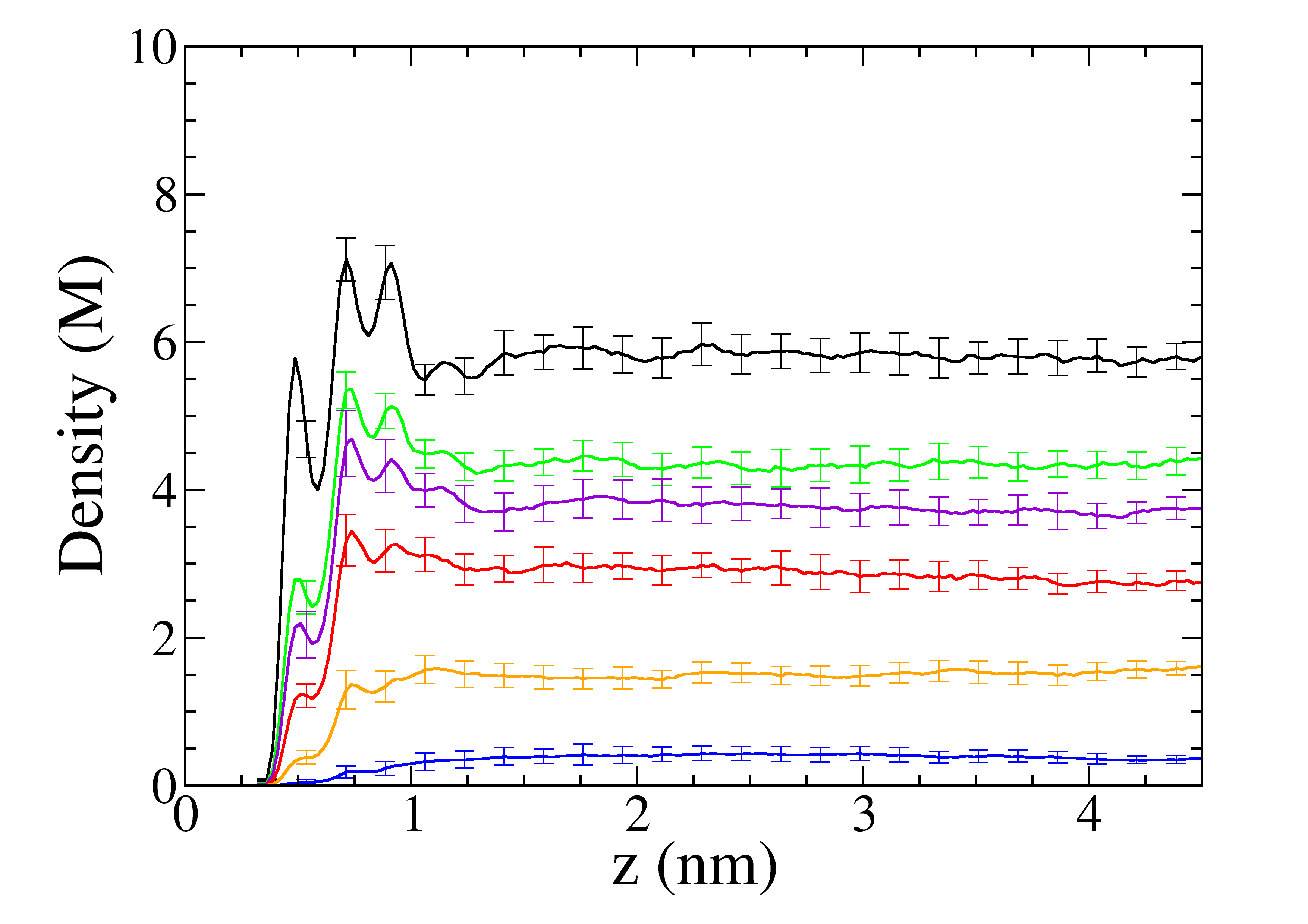}
    \caption{ Cl$^-$ (NaCl)} 
    \label{fig:clna} 
    \vspace{2ex}
  \end{subfigure}%% 
 \begin{subfigure}[b]{0.33\textwidth}
    \centering
   \includegraphics[width=\textwidth]{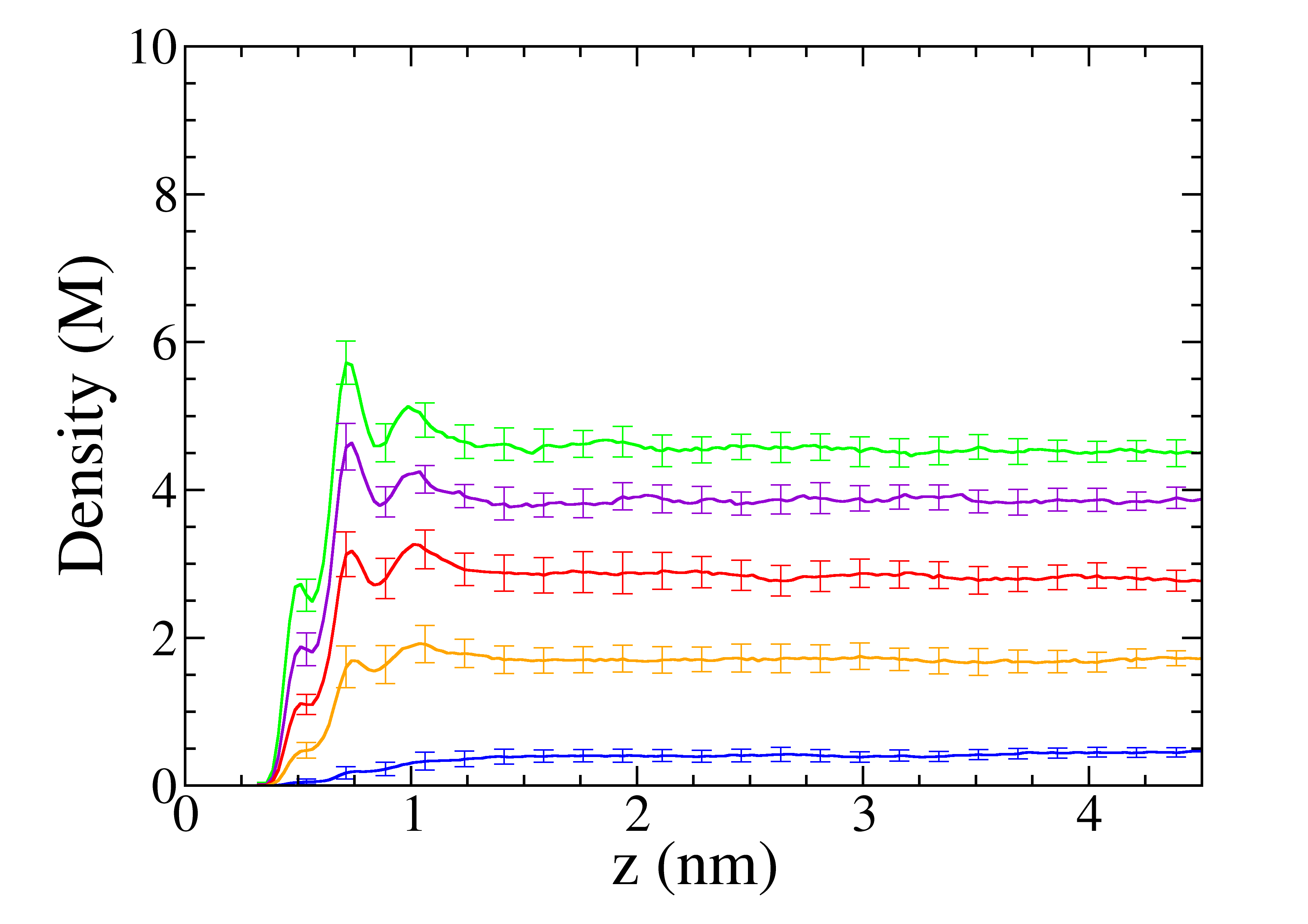}
    \caption{ Cl$^-$ (KCl)} 
     \label{fig:clk} 
    \vspace{2ex}
  \end{subfigure} 
    \begin{subfigure}[b]{0.33\textwidth}
   \centering
    \includegraphics[width=\textwidth]{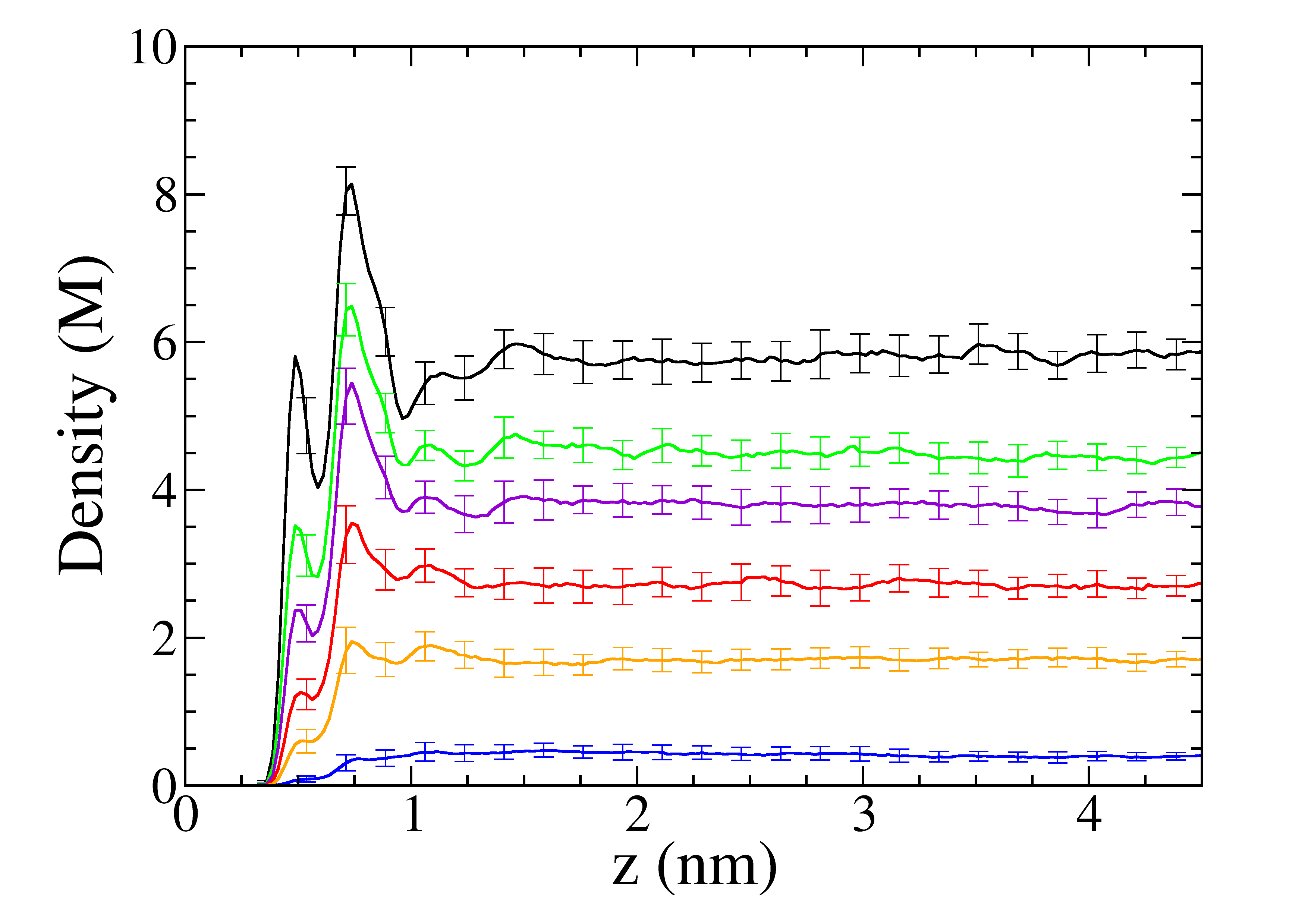}
    \caption{ Cl$^-$ (LiCl)} 
    \label{fig:clli} 
    \vspace{2ex}
  \end{subfigure}%% 
 \caption{  Molar (M) density  of the cations (top row) and the corresponding anions (bottom row) for the three systems considered in this work. Black, green, magenta, red, orange, blue curves correspond to bulk solution concentrations 0.5, 2, 3, 4, 4.4 and 6 M, respectively. }
    \label{fig:numdens}
\end{figure*}

We start this section by reporting in \cref{fig:numdens} the concentration of the different ionic species in solution as a function of the $z$-coordinate, corresponding to the simulation cell direction orthogonal to the surface of negatively charged graphene electrodes. As expected, these profiles show preferential adsorption of cations at the electrode surfaces. For Na$^+$ and Li$^+$, a sharp density peak is observed at a distance of $0.5$ nm from graphene, followed by a second, less pronounced peak at $0.75$ nm. At the highest concentrations, a third cation peak emerges around $1.15$ nm, which is more pronounced for Li$^+$. In contrast, in the case of K$^+$, a small peak at $0.3$~nm is followed by a much larger and relatively diffuse density peak at $0.6$~nm. This is due to specific adsorption of the larger cation at the carbon surface, a small number of which partially dehydrate to directly coordinate to carbon.

The difference in the $z$-density profiles for the different systems is less notable when considering Cl$^-$ with respect to cations. At the lowest bulk concentrations, there is a monotonically increasing density which reaches bulk values around 1.5 nm from the graphene interface. As the concentration rises, further density peaks are observed close to the carbon substrate, determined by the emergence of a multi-layered electrical double-layer structure, consistent with previously reported results  \cite{Finney2021,Elliott2020}. In such double-layer configurations, adjacent solution layers, rich in cations or anions, arise at the interface due to ion crowding (as in the case of the cations that are attracted towards the negatively charged surface of the electrode) and ion correlation (the localized positive excess charge in the closest layers to the electrode, in turn, attracts the anions).

%The thermodynamic driving force for the adsorption of the ions at the electrode is largely unchanged in the three different systems (thanks to the simulation protocols adopted), if one considers this determined by the electrode charge and bulk solution concentration. Nonetheless, there are clear ion (finite-size) effects in \cref{fig:numdens}. 
The results reported in \cref{fig:numdens} are consistent with those of \citep{Elliott2022} with  NaCl(aq) and LiCl(aq) systems displaying, qualitatively, the same solution side double layer structure.
The case of KCl(aq) differs somewhat, with the same position of the two first two peaks for \cref{fig:clk} in both cases, but a different intensity compared with \citep{Elliott2022}. In turn, this intensity difference can be due to the different classical force fields used for water, carbon and ions as well as the use of scaled ionic charges not considered here. Besides these rather minor differences, the results presented here seem to be robust with respect to the chosen classical model.
However, other results in the literature (see \citep{Dovckal2022}) show clear qualitative differences (in particular for the KCl(aq) system where no adsorption is observed), most likely due to the lack of dynamic polarization considered for the graphene electrodes.

\clearpage

\subsection{Electrical Double Layer Properties}
In this section we will derive and analyze the electrical properties of the electrode-electrolyte systems considered in this work.

\paragraph{Electrode Charge Screeninig}
We begin by considering the screening factor \cite{Finney2021}  $f$ defined as:
\begin{equation}\label{eq:scrf}
    f(z) = -\int_{0}^{z} \frac{\rho_{ions}(z^\prime)}{\sigma} \de z^\prime
\end{equation}
where $\sigma$ is the superficial charge of the electrode interface and $\rho_{ions}(z)$ is the density charge of ions only, which is considered a function of just the $z$ coordinate, i.e., it is averaged over the $x$ and $y$ coordinates. 

The screening factor represents the extent to which the electrolyte phase electrically screens the charged interface. When $f$ converges to a value of one, the charge on the electrode is entirely shielded by the electrolyte.  By considering only the ions in the calculations of $f$, we can compare their screening potential to predictions of simple mean field models.
The integration shown in \cref{eq:scrf} and \cref{eq:pot} is performed numerically. Data are first smoothed by applying the Savitzky-Golay \citep{Savitzky1964} finite impulse response smoothing filter of order 3 with a window width of 5 points, implemented in Matlab. The smoothed curves obtained are  then integrated using the trapezoidal rule. Error bars are computed by error propagation through the integration procedure. 

\begin{figure*}
 \begin{subfigure}[b]{0.33\textwidth}
    \centering
   \includegraphics[width=\textwidth]{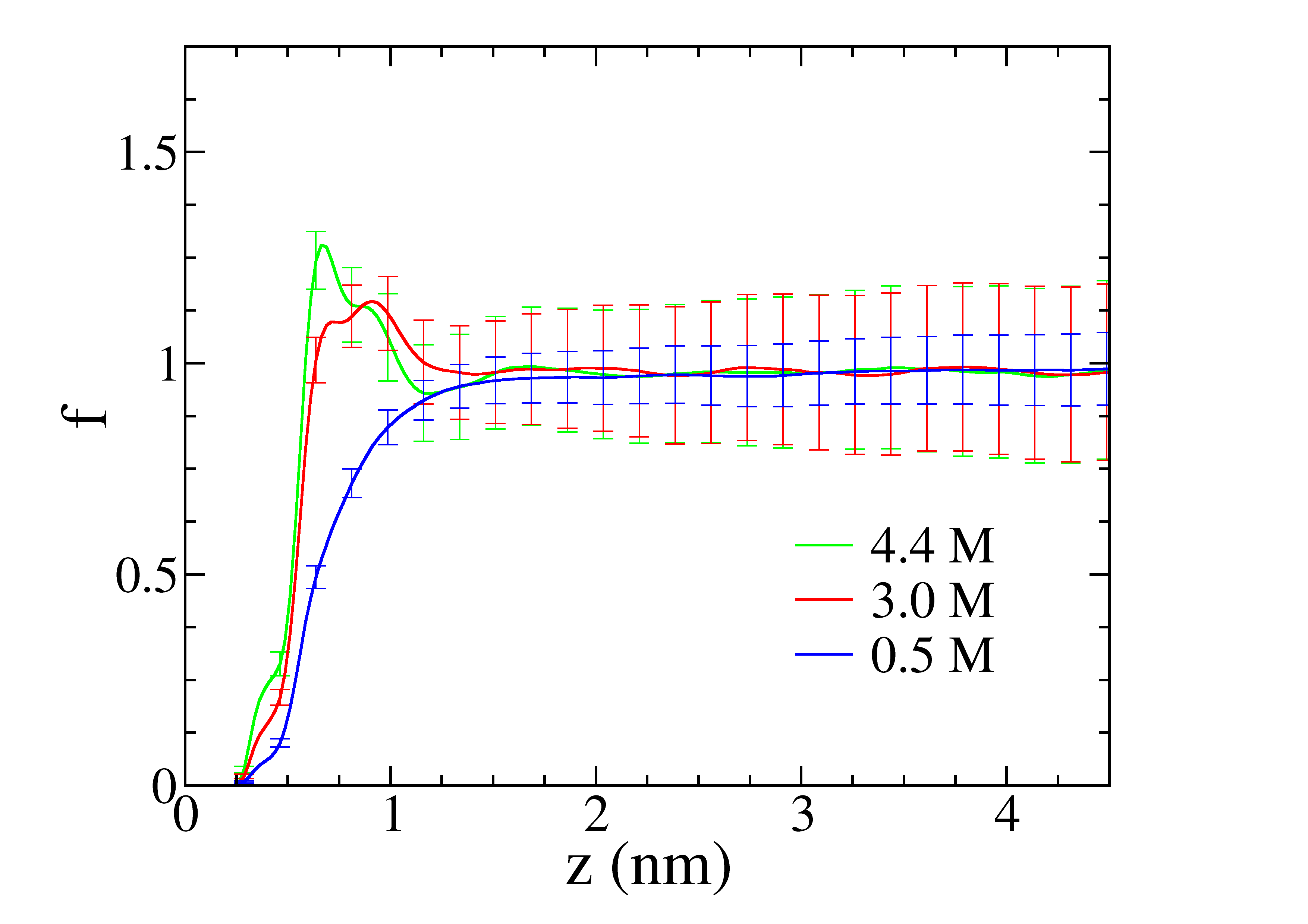}
    \caption{ KCl} 
     \label{fig:fkcl} 
    \vspace{2ex}
  \end{subfigure} 
    \begin{subfigure}[b]{0.33\textwidth}
   \centering
    \includegraphics[width=\textwidth]{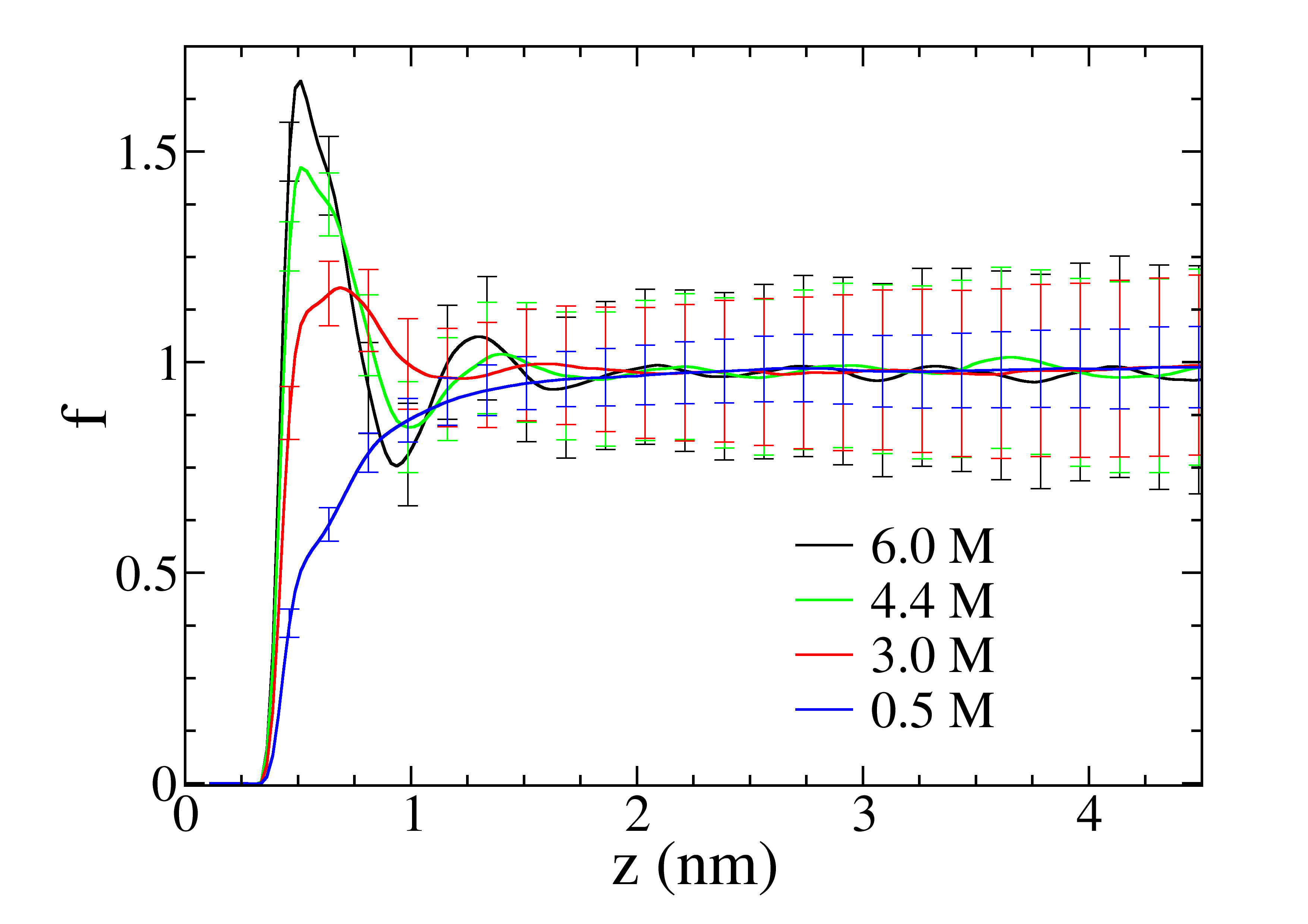}
    \caption{ LiCl} 
    \label{fig:flicl} 
    \vspace{2ex}
  \end{subfigure}%% 
    \begin{subfigure}[b]{0.33\textwidth}
   \centering
    \includegraphics[width=\textwidth]{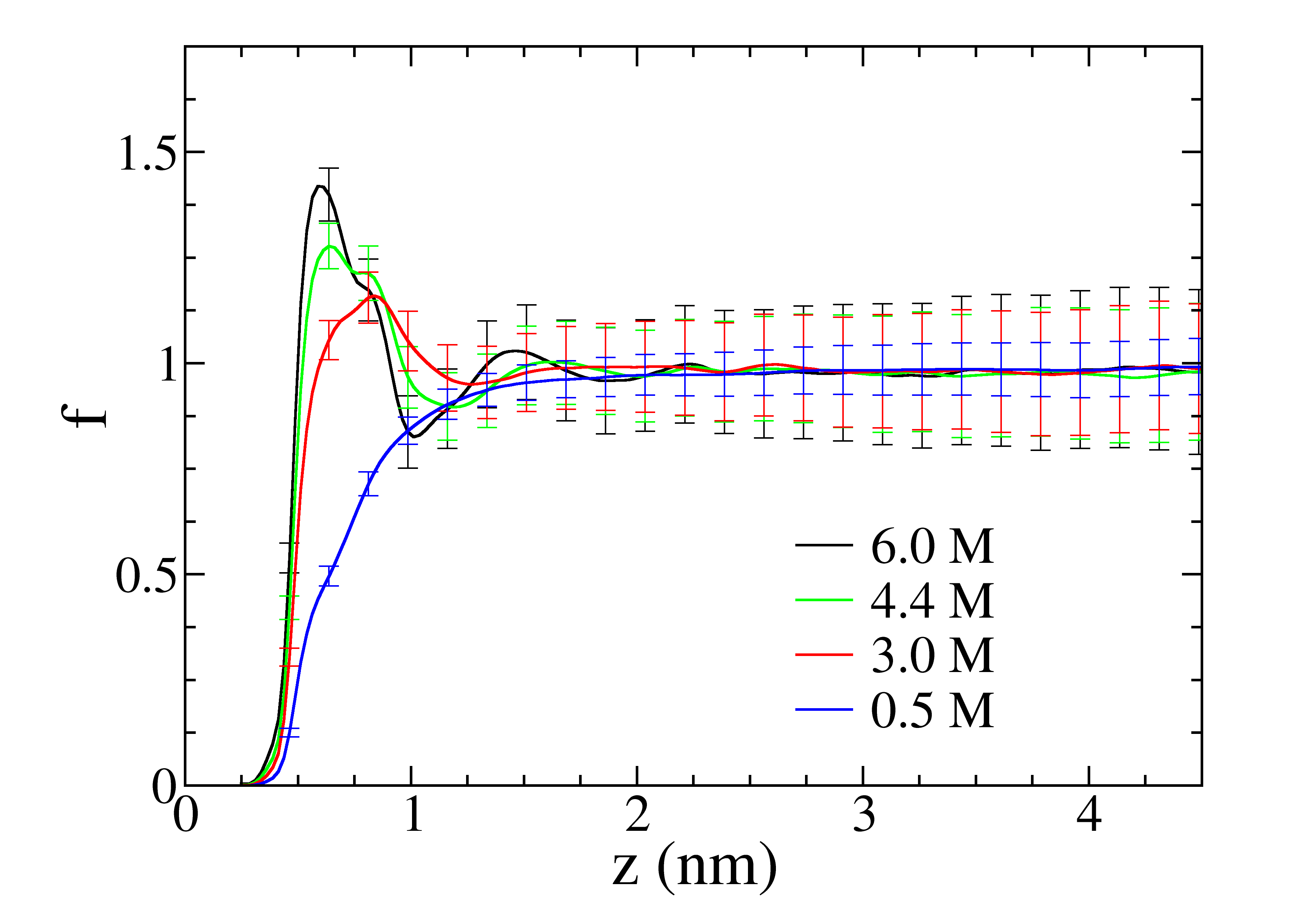}
    \caption{ NaCl} 
    \label{fig:fnacl} 
    \vspace{2ex}
  \end{subfigure}%% 
 \caption{  Screening factor as defined in \cref{eq:scrf} for the three systems considered using ion solution charge densities only. We included only a subset of the concentrations for clarity, and the results for all the concentrations are reported in the SI  (see Fig.S3 of the SI). }
    \label{fig:screenf}
\end{figure*}

The screening factors for all systems are reported in \cref{fig:screenf}. When the concentration of the ions is below 1 M, an under-screening near the interface can be observed. $f$ increases smoothly to a value of one at around $z=2$ nm. This is qualitatively consistent with the predictions of Gouy-Chapman's theory. 
For higher concentrations, however, $f$ transitions to over-screening at relatively small values of the $z$ coordinate. The over-screening, highlighted by the first peak at $z\approx 0.6$ nm reported in \cref{fig:fkcl,fig:flicl,fig:fnacl}, depends both on the particular ion and the bulk concentration. In particular, the LiCl system has the strongest over-screening effect on the electrode across the entire concentration range considered. 
Over-screening is a well-known effect for ionic liquids \citep{Fedorov2008} and is usually not considered important in the electrolyte solutions, as this was only apparent at relatively high concentrations \cite{Elliott2020,Finney2021,Dovckal2022}. The fact that over-screening appears for higher concentration of the solute, in turn, can be linked directly to the structuring of the ions near the interface observed in \cref{fig:numdens}. With the increase in concentration, the density of the cations closest to the electrode increases with respect to their value in the solution bulk  (see \cref{fig:numdens}). The excess charge associated with this ion accumulation is balanced in adjacent solution layers until the average bulk density is reached \citep{Merlet2013}. 
This description is consistent with our observations, where lithium and sodium show a high degree of structuring near the interface relative to potassium (i.e., multiple ion density peaks are observed, accompanied by a significant over-screening effect). In contrast, potassium, with the lowest degree of structuring near the interface, shows the smallest over-screening among the three ion solutions considered.
Moreover, for potassium, we observe a variation in the slope of the screening factor when $z \approx 0.5$ nm, which increases (becoming more pronounced) as a function of concentration. This additional feature in the screening factor, absent in NaCl and LiCl, can be explained by the direct coordination of the K$^+$ (i.e., through the first coordination sphere) to carbon atoms (as also observed in \citep{Elliott2022}), as opposed to the behaviour of the cations in LiCl(aq) and NaCl(aq) systems (see the first peak at $\approx$0.35 nm in \cref{fig:k} with respect to the first peak at  $\approx$0.5 nm in  \cref{fig:li,fig:na}).

%Among the three systems, the one which greatly differs from the others is the KCl(aq). Potassium cations have the tendency to be directly 
%In \citet{Elliott2022}, it was shown that the K$^+$ could partially dehydrate in its first coordination sphere and directly coordinate to atoms at the electrode surface.

%

\begin{figure*}
 \begin{subfigure}[b]{0.5\textwidth}
    \centering
   \includegraphics[width=\textwidth]{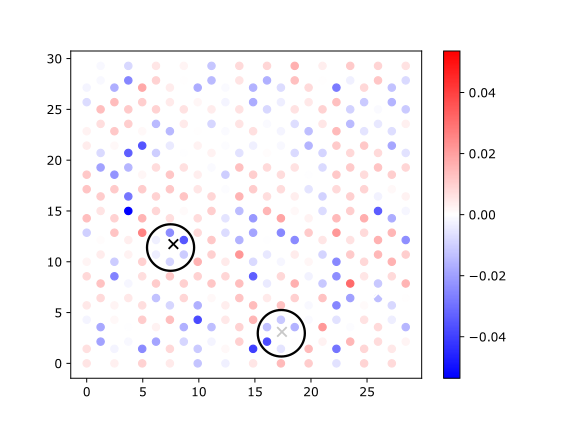}
    \caption{ 0.5 M} 
    \vspace{2ex}
  \end{subfigure} 
   \begin{subfigure}[b]{0.5\textwidth}
   \centering
    \includegraphics[width=\textwidth]{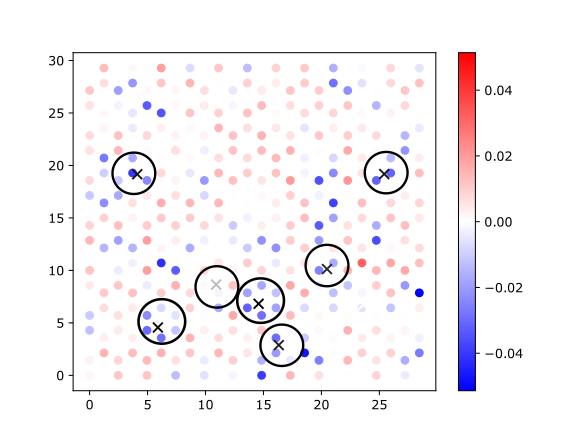}
    \caption{ 4.4 M} 
    \vspace{2ex}
  \end{subfigure}%% 
 \caption{  Representative plot of the computed Mulliken charges on the graphene sheet charged with 4 and in contact with kCl solutions at different concentrations. Circled X’s mark the coordinates of K ions directly adsorbed on the
surface. }
    \label{fig:ads}
\end{figure*}

\paragraph{Electrode Polarisation}
The coordination of the K$^+$ with the carbon atoms on the graphene electrode is shown qualitatively 
in \cref{fig:k} for the lowest (0.5 M) and the highest concentration (4.4 M) considered here. The plots in \cref{fig:ads} represents a single snapshot in the 150 ns long simulation with the highest number of  potassium cations in direct contact with the interface (i.e., at a distance of $0.26$ nm from the interface).
As expected, the number of K$^+$ in direct contact with the interface increases as the bulk concentration of the cations increases, consistent with the observation in \cref{fig:screenf} for the short-distance (from the electrode) behaviour of the screening factor, which increases with concentration.

The accumulation of K$^+$ in the nearby region to the negative electrode (see \cref{fig:k}) results in an increased non-uniformity of the partial charge distribution on the electrode, with higher negative charges located on the carbons closer to the coordinated K$^+$. This, in turn, demonstrates how polarisation effects are important to be considered in systems where direct coordination of electrolytes to the electrode may occur.

 \begin{figure*}
 \begin{subfigure}[b]{0.33\textwidth}
    \centering
   \includegraphics[width=\textwidth]{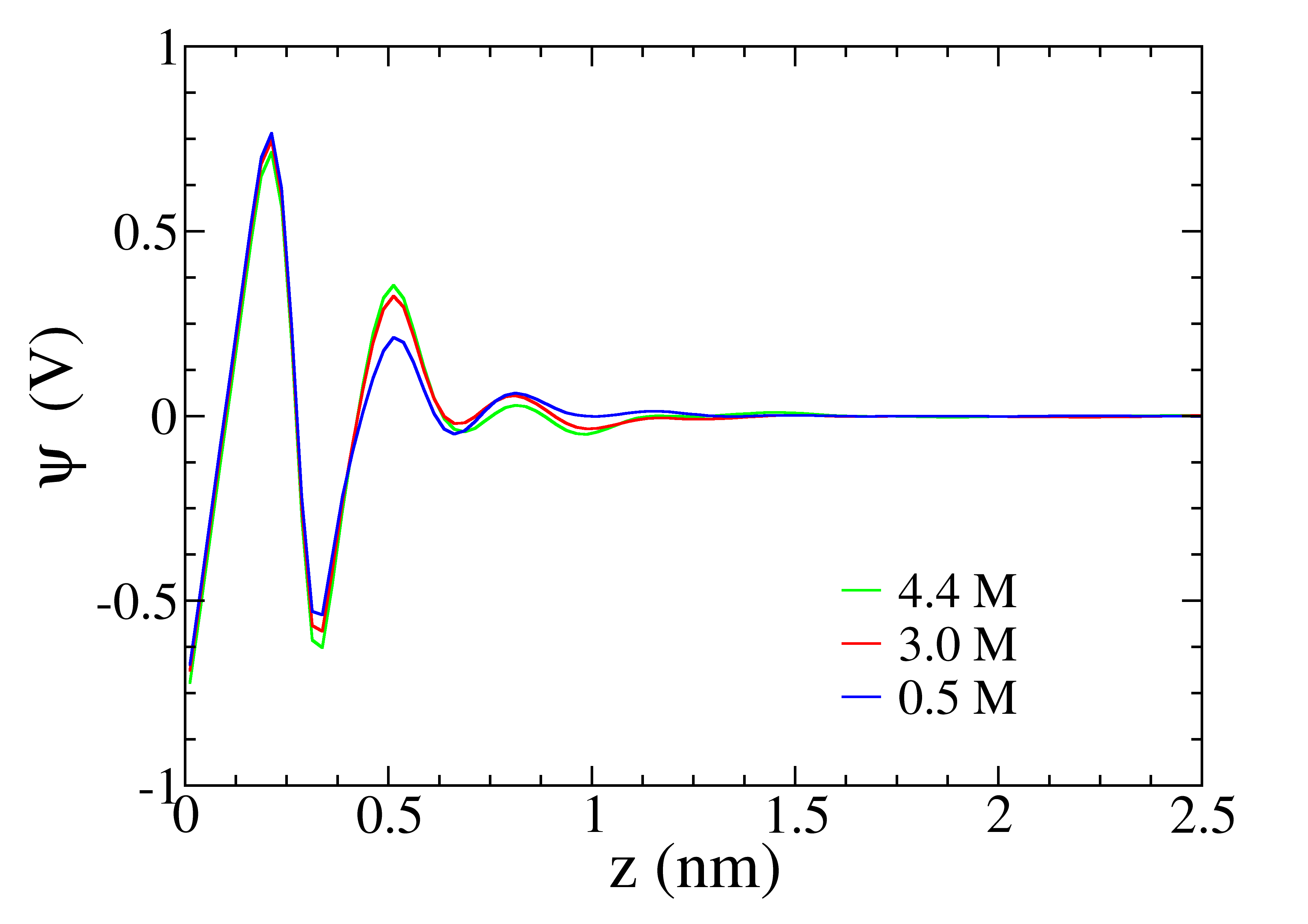}
    \caption{ KCl} 
     \label{fig:Ekcl} 
    \vspace{2ex}
  \end{subfigure} 
    \begin{subfigure}[b]{0.33\textwidth}
   \centering
    \includegraphics[width=\textwidth]{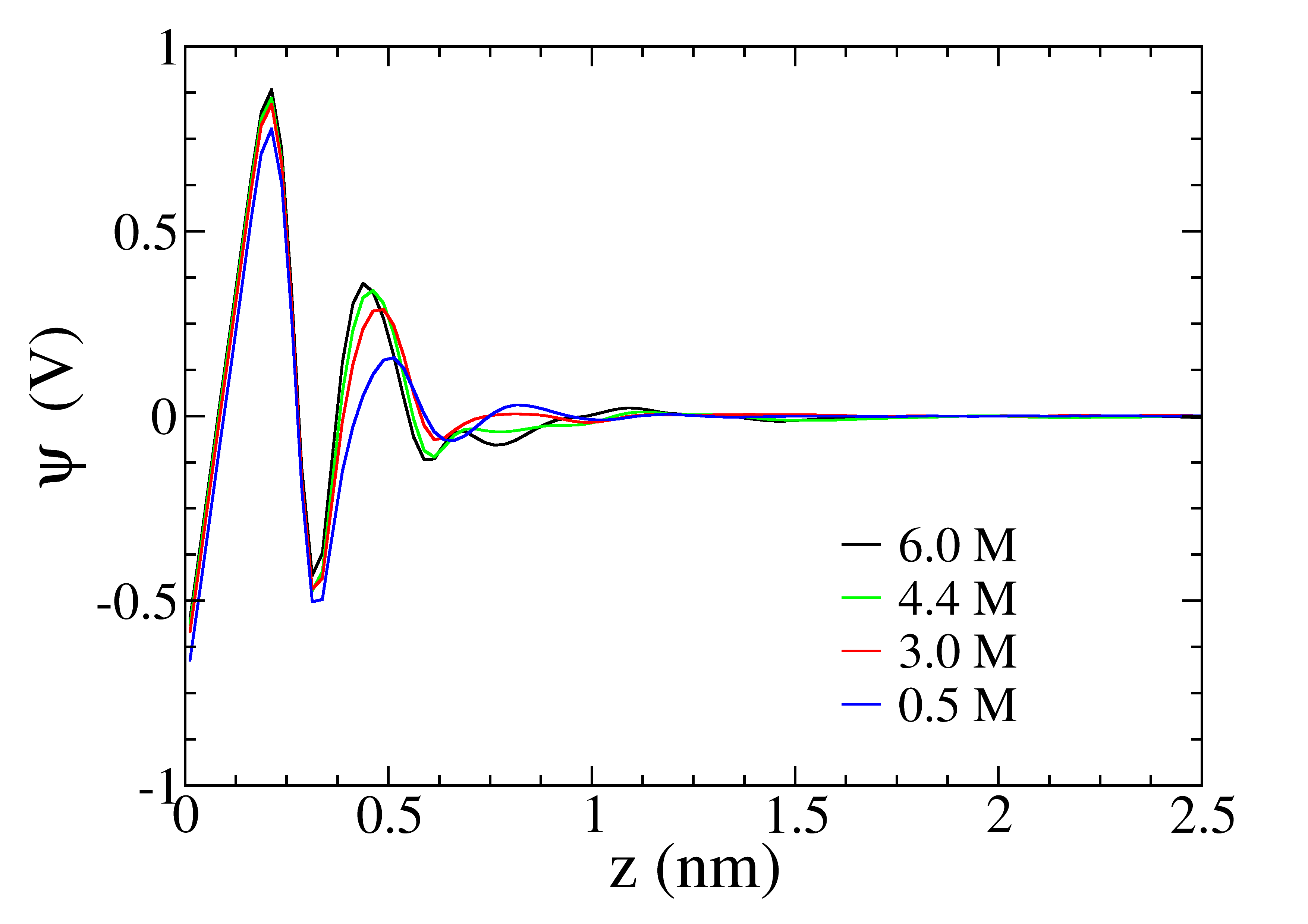}
    \caption{ LiCl} 
    \label{fig:Elicl} 
    \vspace{2ex}
  \end{subfigure}%% 
   \begin{subfigure}[b]{0.33\textwidth}
   \centering
    \includegraphics[width=\textwidth]{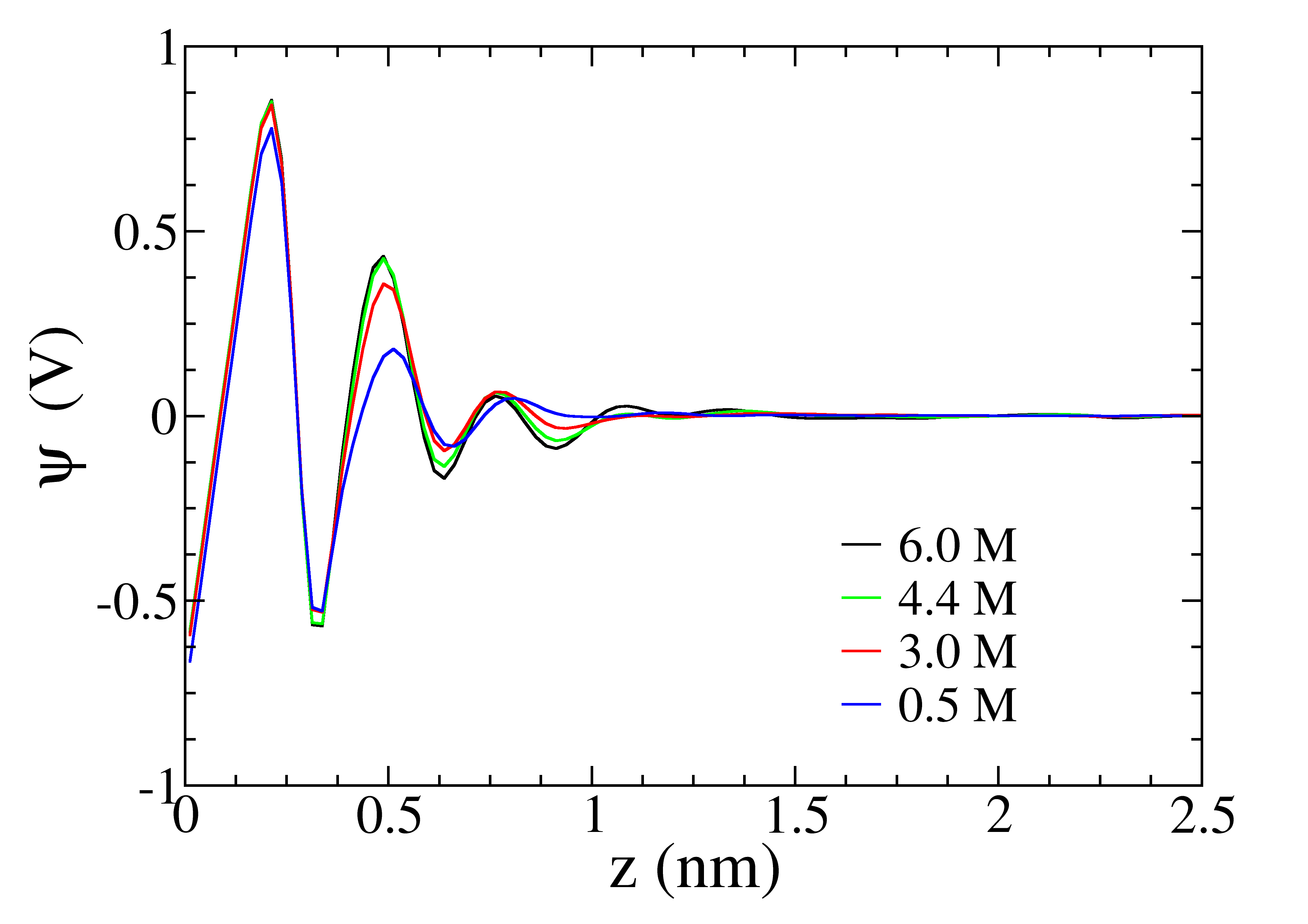}
    \caption{ NaCl} 
    \label{fig:Enacl} 
    \vspace{2ex}
  \end{subfigure}%% 
 \caption{  Electrostatic potential as defined in \cref{eq:pot} for the three systems considered. We included only a subset of the concentrations for clarity. We included the results for all the concentrations in the SI (see Fig.S3 of the SI). }
    \label{fig:potential}
\end{figure*}

%%%%%%%%%%%%%%%%%%%%%%%%%%%%%%%%%%%%%%%%%%%%%%%%%%%%%%%%%%%%%%%%%%%%%%%%%%%%%%

\paragraph{Electrical Potential in the Double Layer}

We calculated the electrical field $E(z)$ and the electrical potential, $\psi(z)$ in the direction orthogonal to the interface using the Poisson equation:%
\begin{equation}\label{eq:poisson}
    -\totnd{\psi(z)}{z}{2} = \totd{E(z)}{z} = \frac{\rho(z)}{\epsilon(z)}
\end{equation}
\noindent where $\rho(z)$ is the charge density calculated for all atoms on the perpendicular axis and we defined $\epsilon(z)=\epsilon_r(z) \epsilon_0$, the product of the permittivity in vacuum $\epsilon_0$ and relative permittivity $\epsilon_r$. It was reported that this latter quantity could be a function of the distance from the electrode \citep{Olivieri2021}, a function of the concentration of the electrolyte \citep{Yang2017}, or possibly both. Given such uncertainties, we consider a constant relative permittivity equal to one in this work.
The electrical potential, $\psi$(z), is obtained from \cref{eq:poisson} by integrating twice with respect to the $z$-coordinate:
\begin{equation}\label{eq:pot}
    \psi(z) = - \int_{0}^{z}\int_{0}^{z^\prime}\frac{\rho(\zeta)}{\epsilon(\zeta)}\de \zeta \de z^\prime
\end{equation}
The two integration constants in \cref{eq:pot} are chosen to set the electrostatic field and potential equal to zero in the bulk, which amounts to considering the bulk as the reference for the calculation of the electrostatic potential.

The results of \cref{eq:pot} are reported in \cref{fig:potential} for a selection of concentrations (see Figure S.3 of the SM for the entire range of concentrations).
In stark contrast to the exponential behaviour predicted by models based on the Gouy-Chapman double layer theory which treats the solvent medium as a continuum with known dielectric, atom/molecule finite-size effects give rise to an undulating $\psi(z)$ function in the interfacial region at all concentrations and in all systems. When calculating the charge distribution, we include all solution species, including water molecules partial charges. Hence, it is unsurprising that the structuring of ions and water molecules at the interface gives rise to a significant departure from the predictions of simple mean field models. Indeed, these finite size effects are a well-reported feature of electrode-electrolyte systems \cite{Kornyshev2007,Snook1981}.

From a relatively large negative value of the potential at the electrode, the (partial) charges of ions and water give rise to fluctuations that attenuate at larger values of $z$, where the bulk solution behaviour is recovered. Generally, increasing the bulk solution concentration increase the amplitude of $\psi (z)$ fluctuations. Furthermore, it is evident from \cref{fig:Elicl} and \cref{fig:Enacl} that the crowding of ions in the double-layer increases with concentration as the positions of peaks and minima in $z$ shift to lower values, a feature also observed by \citet{Finney2021} with graphite and which was related to changes in the screening factor. This concentration dependence is less apparent in the case of KCl(aq), where the value of $\psi(z)$ at the first maximum is less susceptible to changes in the concentration as opposed to NaCl(aq) and LiCl(aq).

\paragraph{Electrical Double Layer Capacitance}

The total capacitance $C_{TOT}$ in these kinds of systems is usually considered as composed of three independent components combined in series: the Electrochemical Double-Layer Capacitance (EDLC), $C_{EDL}$,  and the quantum capacitance (or the space charge capacitance)($C_{Q}$), depending on the spatial distribution of the charges on the graphene \citep{Elliott2022}. 
The total capacitance is then given by 
\begin{equation}
    \frac{1}{C_{TOT}} =  \frac{1}{C_{EDL}} +  \frac{1}{C_{Q}} 
\end{equation}
From \cref{fig:potential} we can easily derive the potential drop, $\Delta \psi$, across the interface as\footnote{ A more precise notation for the potential drop across the interface  would have been $\Delta \Delta \psi = \Delta \psi^- - \Delta \psi_{ref}$. } \citep{Elliott2022} $\Delta \psi = \Delta \psi^--\Delta \psi_{ref}$ where $\Delta \psi^-$ and $\Delta \psi_{ref}$ represent the potential drop at the interface with respect to the bulk for the charged and neutral electrodes, respectively. As a reference for the calculation of the potential drop, we use the potential at the interface in a neutral electrode with all other conditions unchanged. We report the calculation of the potential across the system for a neutral electrode in the SI (see Figure S.4 of the SI) along with the potential drop at the interface ($\Delta \psi_{ref}$) (see Table S.2 of the SI). 
With this definition of the potential drop, the EDLC can be obtained, as  
\begin{equation}
    C_{EDL} = \frac{\sigma}{\Delta \psi }
\end{equation}

The quantum capacitance instead is obtained by calculating the differential quantum capacitance $C_{Q}^{diff}$ according to \citep{Elliott2022}:
\begin{equation}
    C_{Q}^{diff}(\psi) = \frac{e^2}{4\kb T}\int_{-\infty}^{\infty} \sqp{D(E)\mbox{sech}^2\cip{E+\psi}}\de E
\end{equation}
Where $e$ is the electron charge, $E$ is the energy relative to the Fermi level, $D(E)$ is the density of states at a given energy, $\kb$ is the Boltzmann constant, and $T$ is the temperature.
By integrating the differential quantum capacitance with respect to the potential $\psi$ up to the potential drop $\Delta \psi$ calculated for each system, we obtain the integral quantum capacitance $C_Q$:
\begin{equation}
    C_Q = \frac{1}{\Delta\psi}\int_{0}^{\Delta\psi}C_Q^{diff}(\psi) \de \psi
\end{equation}
For more detailed information about the calculation of the quantum capacitance we refer the reader to our previous work \citep{Elliott2022}.

\begin{table}[htp]
\begin{center}
\begin{tabular}{c| c c c c }
        \midrule
   concentration    &  $\Delta \psi$ & $C_{EDL}$ &    $C_{Q}$  &  $C_{TOT}$ \\
        \midrule
         & \multicolumn{4}{ c }{KCl} \\
        \midrule        
0.5 & -1.03  & 6.95 & 10.56 &  4.19 \\
2.0 & -1.01  & 7.10 & 10.31 &  4.20 \\
3.0 & -1.00  & 7.16 & 10.23 &  4.21 \\
4.0 & -0.984 & 7.28 & 9.81  &  4.18 \\
4.4 & -0.988 & 7.25 & 10.07 &  4.22  \\
        \midrule  
 & \multicolumn{4}{c}{LiCl}  \\
         \midrule  
0.5 & -1.05  & 6.82 & 10.78 & 4.18 \\
2.0 & -1.01  & 7.09 & 10.31 & 4.20 \\
3.0 & -1.00  & 7.16 & 10.21 & 4.21 \\
4.0 & -1.00  & 7.16 & 10.21 & 4.21 \\
4.4 & -1.09  & 6.57 & 11.25 & 4.15 \\
6.0 & -1.02  & 7.02 & 10.44 & 4.20 \\
        \midrule  
& \multicolumn{4}{c}{NaCl} \\
        \midrule  
0.5 & -1.05  & 6.82 & 10.78 & 4.18 \\
2.0 & -1.02  & 7.02 & 10.44 & 4.20 \\
3.0 & -0.997 & 7.18 & 10.17 & 4.21 \\
4.0 & -0.996 & 7.19 & 10.15 & 4.21 \\
4.4 & -0.986 & 7.26 & 10.05 & 4.22 \\
6.0 & -1.00  & 7.16 & 10.21 & 4.21 \\
  \end{tabular} 
  \end{center}
  \caption{Electrostatic potential drop ($\Delta \psi$) across the interface (in V), Electrochemical double layer capacitance $C_{EDL}$, Quantum Capacitance $C_Q$, and total capacitance $C_{TOT}$ (in $\mu$F cm$^{-2}$) for each concentration considered (in M). }
\label{tab:cedl}
\end{table}%

%Our main goal is try to explain this observed behaviour, which is made possible by CmuQMMD framework, with which we can describe at the same time the polarization at the interface, and an open boundary system with a prescribed concentration of the ions. 

The results for $C_Q$, $C_{EDL}$, and $C_{TOT}$ for all of the systems considered are reported in \cref{tab:cedl}.
The data show that the total capacitance is practically constant across all the concentration range and for all solution types. The largest variation in $C_{TOT}$ we obtained among all the systems is $\approx$2\% (between the LiCl(aq) and KCl(aq) at 4.4 M).
This result contrasts with the different behaviour of the three cations in solution and near the electrode interfaces, as highlighted in the discussion of the number density of ionic species at the interface (see \cref{fig:numdens}) their screening effect on the charge of the electrode (\cref{fig:screenf}), and as further discussed in the following section in relation to their clustering properties. 

An important point we want to highlight here is that such differences in the behaviour of the cation in solution can be correctly captured through the use of a simulation protocol that combines the pseudo-open boundary condition, i.e.,  C$\mu$MD to maintain constant composition electroneutral bulk solutions beyond the double layer, and the quantum mechanical description for the distribution of partial charges of the electrode.
However, while the capacitance is a critical parameter for this kind of system's applications as supercapacitors, we showed here that the physics of the interfaces between graphene electrodes and electrolytes is much richer than the one captured by such quantity.

\subsection{Ion Association}

An often overlooked effect in systems in alkali chloride solutions is the tendency for ions to associate, forming clusters. In particular, even simple salt solutions exhibit significant non-ideal behaviour at high concentrations. Recent experiments \cite{Hwang2021} and simulations \cite{Finney2022} have shown that extended liquid-like clusters exist in bulk NaCl(aq) at high concentrations and the extent of these ionic networks is promoted in the double layer at carbon surfaces \cite{Finney2021}.  
Since the effectiveness of the graphene-electrolyte devices often depends on the ability to `build up the double layer' (i.e., accumulate ions from the bulk solution in the interfacial region), the structure and mobility of ion species can be essential to this.

%In our work we will use the work cluster to identify an aggregate of atoms of a certain type (in this case cation and anions) which fulfill certain geometric constraints which we will shortly discuss. In turn, the definition we are using to identify a cluster implies that we are considering as a cluster any aggregate of cation and anions of any dimension, appearing in the system, independently of its life-time. 

%In order to define the average contact distance between the ions, we calculated the cation-anion RDF for all the different pairs we considered (see Fig. S.1 of the Supplementary Material). We considered as clustering distance the value of $r$ corresponding to the first minimum of the RDF:  0.39 \AA~for KCl, 0.29 \AA~for LiCl (\hl{should we clarify that for Li the first peak is very small?}) and 0.34 \AA~for NaCl.
\paragraph{Ion Clusters}
To identify and characterise ion associates in the simulations in this work, pairwise RDFs were computed (see Fig. S.1 of the Supplementary Material, (SM)), and the first minima in these informed truncation distances ($r_c$) for first-sphere ion-ion coordination. $r_c=0.29$, 0.34 and 0.39 nm for Li-- Na-- and K--Cl, respectively, reflecting the different sizes of the cations. Clusters were identified as fully connected networks in the graph of adjacent ion-ion connections according to this geometric criteria, regardless of their total charge or lifetime.

\begin{figure*}
  \begin{subfigure}[b]{0.5\textwidth}
   \centering
    \includegraphics[width=\textwidth]{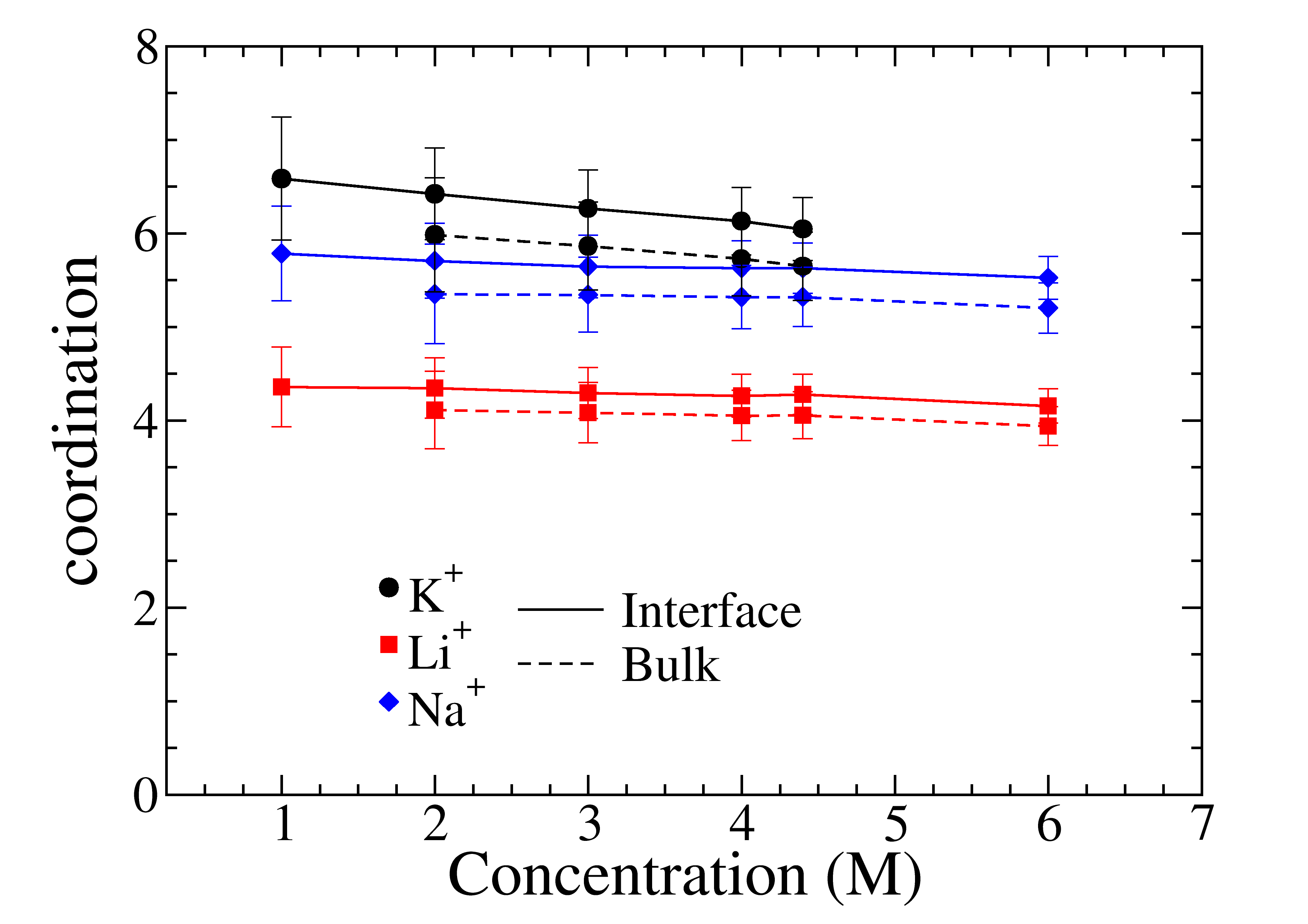}
    \caption{ Cation/Water} 
    \label{fig:fcatw} 
    \vspace{2ex}
  \end{subfigure}%% 
 \begin{subfigure}[b]{0.5\textwidth}
    \centering
   \includegraphics[width=\textwidth]{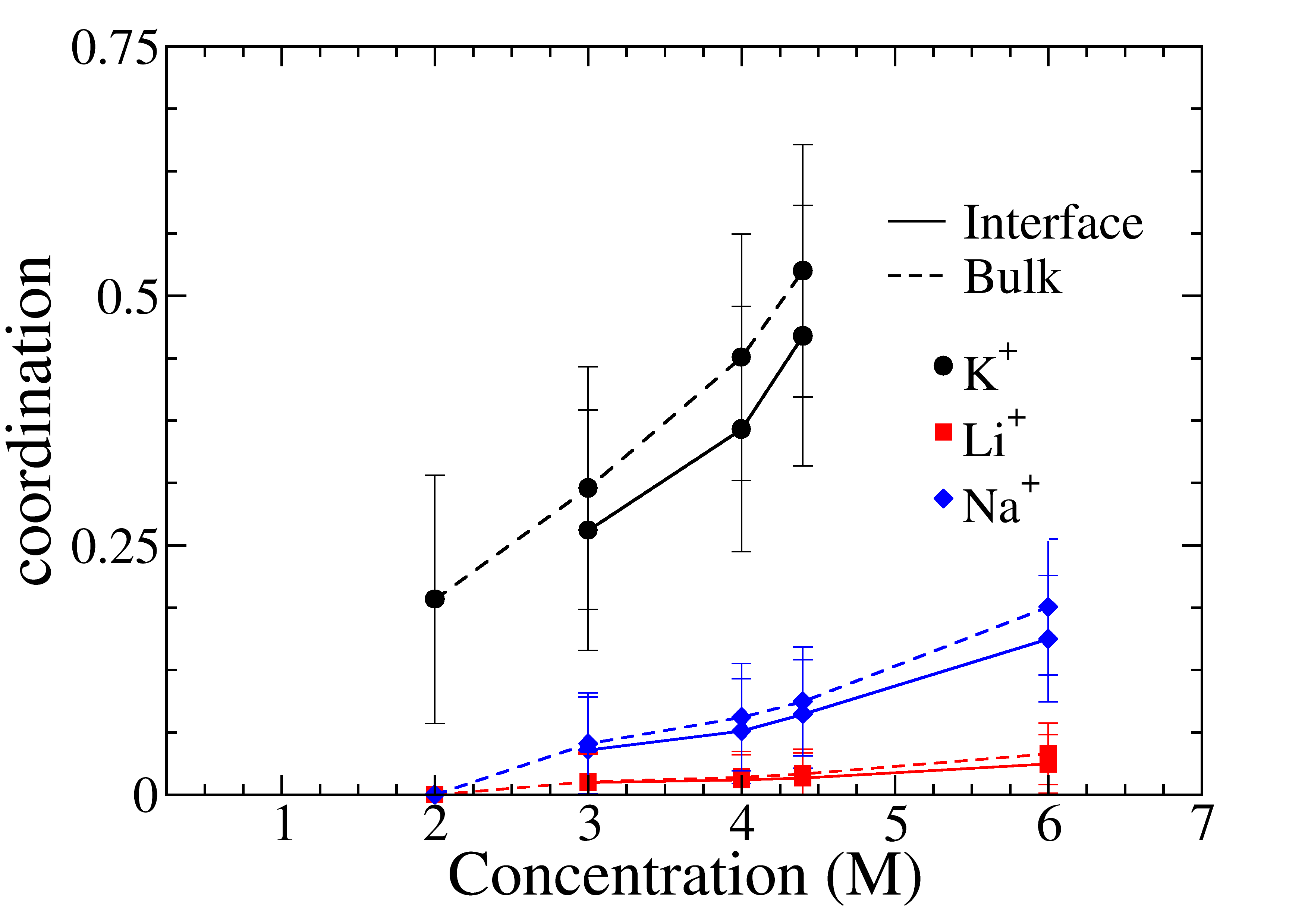}
    \caption{ Cation/Anion } 
     \label{fig:fcatan} 
    \vspace{2ex}
  \end{subfigure} 
 \caption{  Coordination number for the different systems at the different concentrations. }
    \label{fig:contact}
\end{figure*}

\Cref{fig:contact} provides the average first-sphere coordination number between cations and O of water (see \cref{fig:fcatw}) as well as cations and anions for all systems, calculated using \cref{eq:sphere}. 

%The coordination number represents the ensemble average of the number of contacts (according to the geometric rule just defined) between ions in the system. Therefore, a large coordination number implies either a large number of smaller cluster or few big structures, or both. In any case, the higher is the coordination number, the higher is the tendency of ions to form clusters.
The results shown in \cref{fig:contact} indicate no significant surface effect on the coordination of cations with water or chloride when ions in the interface ($0<z<2.5$ nm) and bulk ($2.5<z<4.5$ nm) regions were investigated. There is a slight increase in the mean cation-anion coordination, and a concomitant decrease in cation-water coordination, at the interface compared to the bulk; however, this difference is within the margin of error.
Generally, the effect of increasing concentration is to increase the number of cation-anion contacts, particularly for KCl(aq), where the coordination number is more than double that of the other systems for all concentrations (and with Li--Cl coordination being negligible even at 6 M). From the largest to smallest variation in the coordination number we can write $\mbox{K}^+ \rightarrow \mbox{Na}^+\rightarrow \mbox{Li}^+$. This trend follows the decrease of the ion radius and it is likely due to the stronger binding of water in the solvation spheres of smaller cations. Furthermore, the average cation-water coordination number is unchanging with a concentration within the margin of error.

%The first result we obtained is that the graphene interface does not have a large impact on the behaviour of the ions in solution in all the concentration range.
%The larger effect is obtained for potassium, but even in this case, the differences between the coordination number in the bulk and interface is within the errors.
%Another effect we obtain is the variation of the coordination number with respect the concentration. While the coordination between cations and water molecules can be considered with a good approximation flat in all the range of concentrations, the Cation/Anion contact exhibits a variation over the concentration depending on the cation type in this order, from the largest to the smallest variation: $\mbox{K}^+ \rightarrow \mbox{Na}^+\rightarrow \mbox{Li}^+$. 
%In terms of the magnitude of the coordination number, we can observe that Li does show a negligible coordination number, even at the highest concentration we considered (i.e., $6$M). 

In simulations of NaCl(aq) in contact with graphite, \cite{Finney2021} the substrate was found to increase cation-anion correlations in the double layer with respect to the bulk, particularly beyond $5$ M. It is important to note that different models (due to the different system) were used and also that system size likely plays a role in the extent that clusters can grow (both in e.g., the system-size dependence of the availability of ions to form associates and the extent to which finite-size and percolating clusters may form in effectively confined canonical systems.)

\begin{figure*}
 \begin{subfigure}[b]{0.5\textwidth}
    \centering
   \includegraphics[width=0.95\textwidth]{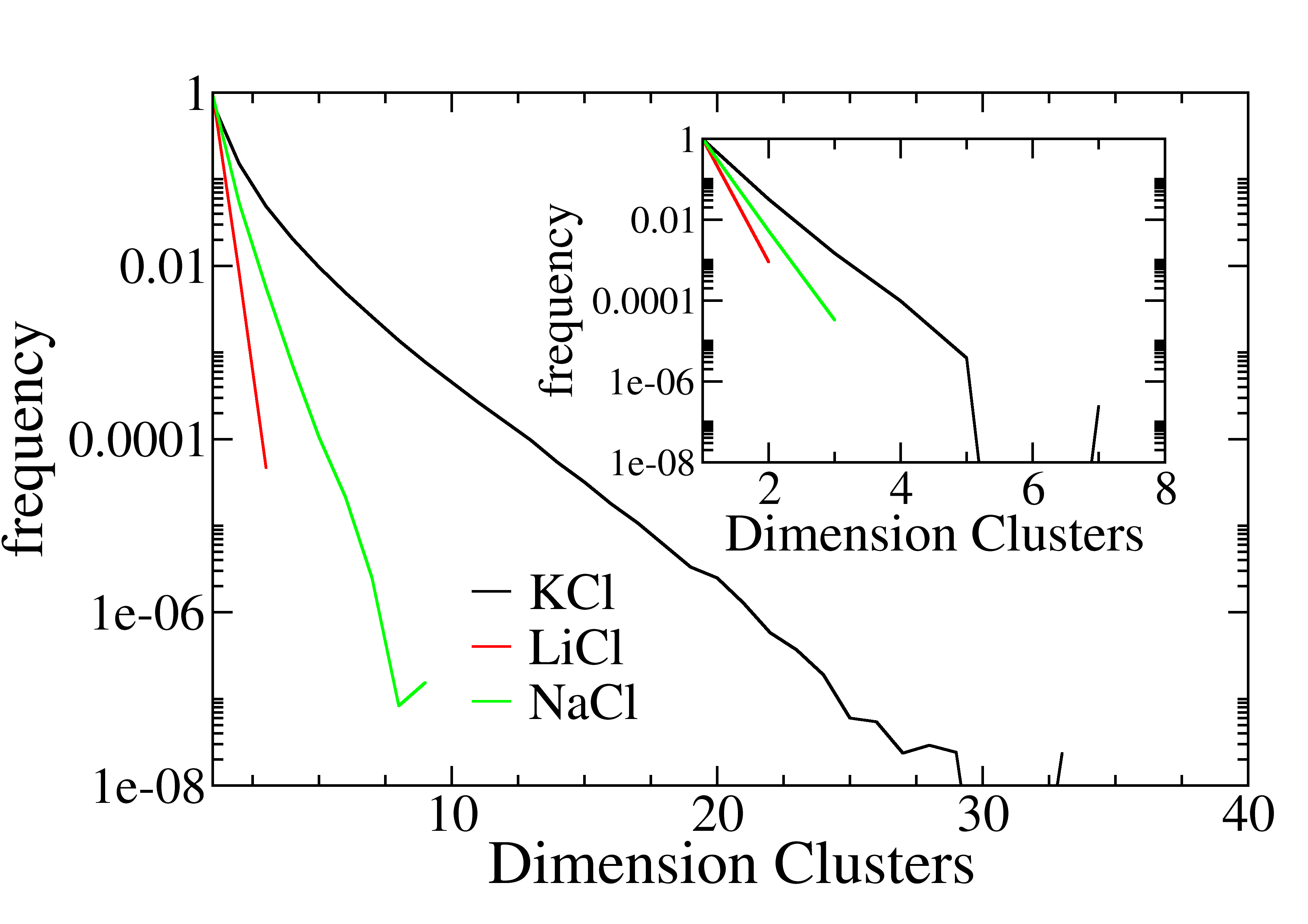}
    \label{fig:subhisto} 
       \vspace{4ex}
  \end{subfigure} 
  \begin{subfigure}[b]{0.5\textwidth}
    \centering
   \includegraphics[width=0.9\textwidth]{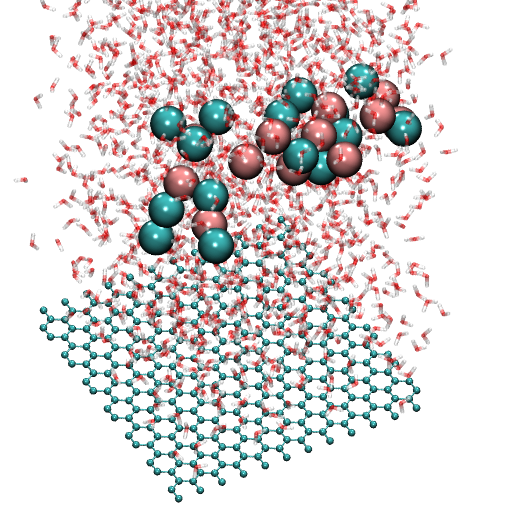}
%   \label{fig:subclust}
  \end{subfigure} 
 \caption{ On the left: Histogram of the relative frequency of the cluster of different sizes for the concentration of 4.4 M. In the inset, the same quantity for the 0.5 M case. On the right: an example of a cluster composed of 26 ions for the KCl system at 4.4 M. }
    \label{fig:histo}
\end{figure*}

The change in coordination for different salts is reflected in the cluster size probability distributions presented in \cref{fig:histo} for the case of 4.4 M (we report the results for the entire range of concentrations in Figure S.5 of the SM).
There is a clear difference in the extent to which clusters can grow, with lithium forming clusters containing at most four ions and potassium forming much larger networks containing as many as 35 ions. Even at the highest concentrations, the majority of the Li$^+$ are dispersed in solution, fully solvated in their first shell. A snapshot of a configuration obtained during the simulation of KCl at 4.4 M is shown in \cref{fig:histo}. Although the most probable clusters contain only a few ions (for clusters composed of five ion units, we obtained a relative frequency of ~0.01), larger species do contribute to the charge storage capacity and must be considered. 
What we observe is a stronger tendency of the potassium to associate into large aggregates---albeit ones which are highly dynamic on the timescales of the simulations---compared to sodium or lithium. %can help explain the results in \cref{tab:cedl}, where the $C_{EDL}$ for potassium seems to weakly decrease over the range of concentrations. The structuring of the ions increases for the KCl system with the increase of the concentration.
%, making the system less and less ideal and reducing at the same time the mobility of the ions within the system, which cause a reduction of the capacitance.

\begin{figure*}
  \begin{subfigure}[b]{\textwidth}
   \centering
    \includegraphics[width=0.75\textwidth]{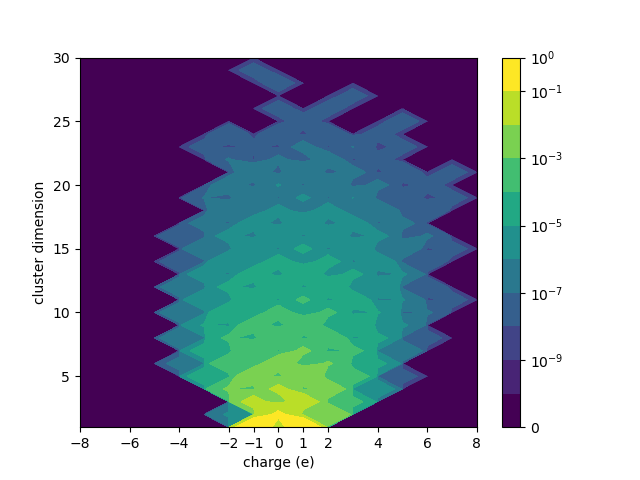}
  \end{subfigure}%% 
 \caption{  2-dimensional histogram (charge VS dimension of the clusters) for the KCl(aq) system at the largest concentration considered (4.4 M). }
    \label{fig:2Dhisto}
\end{figure*}

Since the KCl(aq) system shows the formation of large aggregates of ions, it is interesting to study the relative frequency of the charge of these aggregates. In \cref{fig:2Dhisto} we plot the 2-dimensional histogram showing the relative frequencies of the charge vs. the cluster size for the KCl(aq) system. 
The histogram is skewed towards positive charges, with the appearance of clusters containing an excess of positive charge as large as +7e, although the majority of the clusters are neutral. %The fact that the clusters capture the cations more than anions %\hl{was already observed in X} 
%shows that for highly concentrated solutions of systems with a high tendency to form aggregates (such as the K$^+$ in KCl(aq) systems), the charge mobility necessary to form a double layer  can be reduced even more by the formation of such aggregates. 
%

\paragraph{Ion Mobilities}
As well as a high capacity to store charge, an optimal charge storage device must also be a good electrical conductor. In this regard, we might expect the conductivity of solutions to decrease when clusters are present. Indeed, this can be perceived as a relative decrease in the activity of charge carriers due to increasingly non-ideal solutions. To test this, we calculated the conductivity of bulk NaCl(aq) solutions with concentrations ranging from 1--10 M from the ion diffusion coefficients calculated by Finney and Salvalaglio in finite size systems and in the dilute limit \cite{Finney2022,Finney2021Bridg}. 
To determine conductivity we use the Nernst-Einstein equation:
\begin{equation}
    \sigma_{NE} = \frac{e^2}{V k_\mathrm{B} T} (N_+z_+^2D_+ + N_-z_-^2D_-)
\end{equation}
where $e$, $V$, $k_\mathrm{B}$ and $T$ are the elementary charge, simulation cell volume, Boltzmann's constant and temperature, respectively. 
$N$ and $D$ are the total number of ions and diffusion coefficients for ions with charge indicated by the subscript. Furthermore, we assume that, given the highly dynamic nature of the clusters observed in solution, the valency of ionic species, $z$, is equal to one.

\begin{figure}[ht]
    \centering
    \includegraphics[width=.55\textwidth]{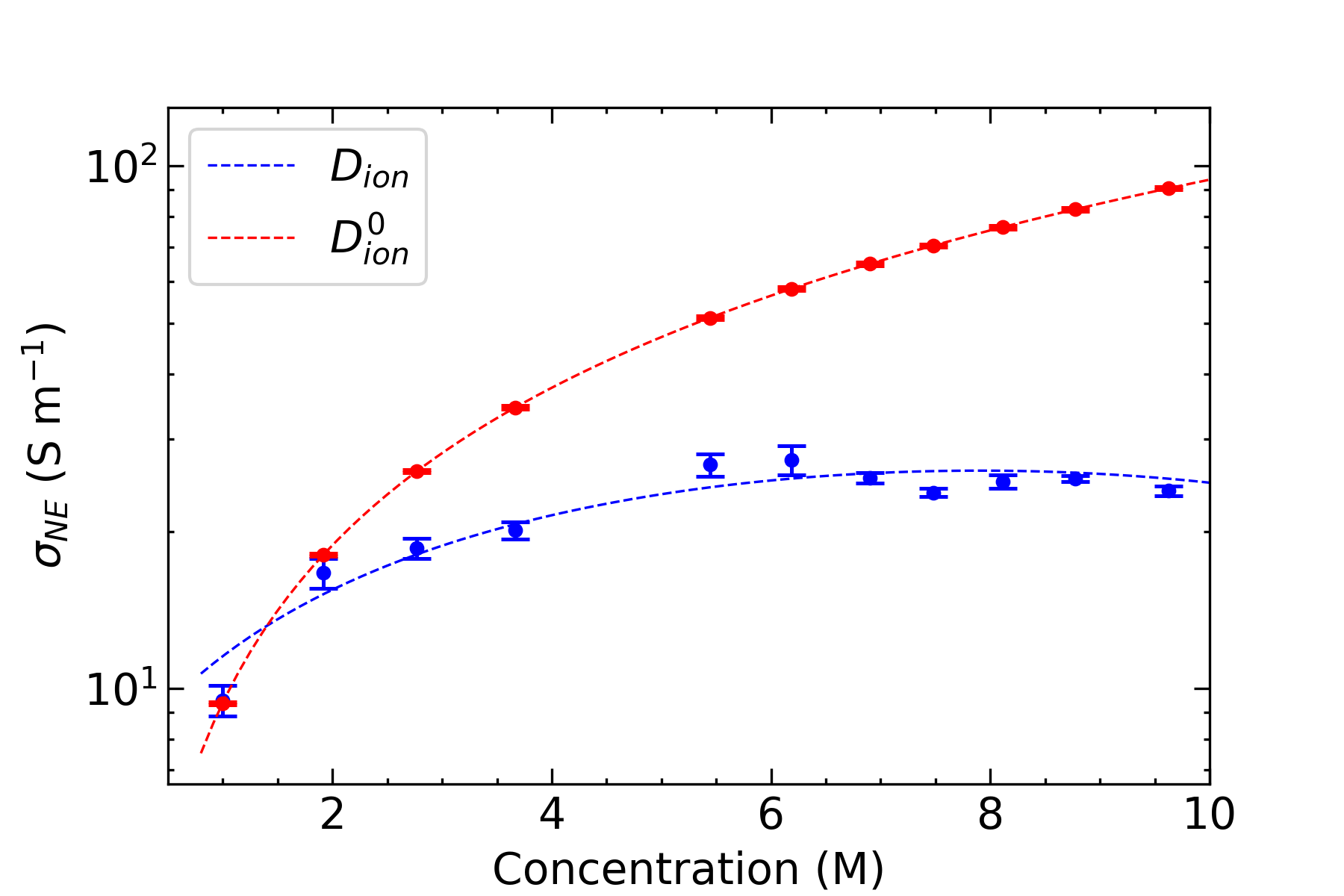}
    \caption{Solution conductivities, $\sigma_{NE}$, of bulk NaCl(aq) solutions calculated for a range of concentrations. To this aim, the Nernst-Einstein equation was adopted where ion diffusion coefficients were determined from simulations at finite concentration, $D_{ion}$ (blue), or from a single simulation at the dilute limit, $D_{ion}^0$ (red). Dashed lines are a guide for the eye, while error bars indicate uncertainties in the conductivities associated with the calculated $D$ value from Refs. \cite{Finney2022} and \cite{Finney2021Bridg}. %\hl{See alternative representation in Figures\slash conductivity-ratio.png}
    }
    \label{fig:conduct}
\end{figure}

\Cref{fig:conduct} provides the solution conductivities for NaCl(aq) where either the diffusion of ions in finite concentration simulations was used ($D_{ion}$) or the diffusion of ions in the dilute limit ($D_{ion}^0$) was considered. For the latter, ions are assumed to be completely, dispersed as association, even beyond the second solvation sphere, did not occur in simulations at the dilute limit.
For the estimate of $D_{ion}^0$, \citet{Finney2021Bridg} performed extended simulations of a single cation and anion in 4,000 water molecules; here, $D_+^0 = 1.223 \pm 0.005 \times 10^{-5}$ cm$^2$ s$^{-1}$ and $D_-^0 = 1.282 \pm 0.008 \times 10^{-5}$ cm$^2$ s$^{-1}$. In all cases, diffusion coefficients were corrected to account for simulation finite size effects \citep{Yeh2004}. Unsurprisingly, a linear correlation in $\sigma_{NE}$ as a function of concentration is found when a constant $D_{ion}^0$ is used for the diffusion of ions, independent of concentration. This is inaccurate at relatively high concentrations, given the simulation and experimental observations of ion-ion correlations\cite{Finney2022,Hwang2021}. 

When accounting for the non-idealities in the solution and the formation of clusters explicitly in the diffusion of ions, we find that the solution conductivity reaches an upper limit between 4 and 5 M. At the lowest concentrations (1--2 M), the conductivity from finite concentration and dilute simulation data agree, and the simulation predictions match well with experimental measurements \citep{Widodo2018}. A crossover in the conductivity behaviour from the `pseudo-ideal' to non-ideal regime occurs between 2 and 3 M. Therefore, over a wide concentration range up to the salt solubility, non-idealities will affect the performance of electrical devices; depending upon the chosen application, electrolytes should be chosen to minimize these effects.

\clearpage

\section{Conclusions}

In this work, we presented an extended set of simulations describing the interface between three different electrolyte solutions - (KCl(aq), LiCl(aq), and NaCl(aq)) - in contact with the surface of a negatively charged graphene electrode. 
To investigate these systems, we combined QM/MD and C$\mu$MD methodologies into a new simulation framework. QM/MD models of the graphene electrode in contact with an electrolyte enabled the explicit coupling of the electrode polarizability with the instantaneous configuration of the electrolyte. The latter was maintained in equilibrium with a liquid phase at constant bulk concentration thanks to the C$\mu$MD model, which mimics open-boundary conditions.

We performed a thorough analysis of the interaction of the ions with the electrode by showing the different behaviour of the three cations in the double layer, focusing on K$^+$, which, according to our results, is able to directly adsorb at the electrode surface at shorter distances compared to Li$^+$ and Na$^+$, modifying the screening effect of the solution.

Calculations of the integral capacitance indicated no concentration dependence or specific ion effects, with a total capacitance of around 4.2 \textmu F cm$^{-2}$ across all systems. However, the lack of variation in capacitance hides the rich electrolyte solution behaviour, particularly for the ions close to the electrode.  
We showed, for example, that large KCl clusters emerge in solution, which might be important when considering properties associated with ion mobility and charge transfer. 

Our results indicate that accurate models of the interface - able to account for the position-dependent non-ideality of  electrolyte solutions - better capture the configurational and dynamical details underpinning the electrochemical behavior of interfaces at the atomistic level, and that is often overshadowed by the calculation of aggregated quantities such as the integral capacitance. We plan to extend our calculations to include a range of charged electrodes, both positive and negative, and further investigate ion dynamics in solution.

\section{Acknowledgements}

We acknowledge the support provided by the IT Services use of the Computational Shared Facility (CSF) and at the University of Manchester. NDP, JDE and PC thank the European Union's Horizon 2020 research and innovation programme project VIMMP under Grant
Agreement No. 760907. ARF and MS acknowledge funding from an EPSRC Programme Grant (Grant EP/R018820/1), which funds the Crystallisation in the Real World consortium.

\bibliographystyle{unsrtnat}
\bibliography{bibliography_20Feb20}

\end{document}